\documentstyle[prd,aps,epsfig,psfig]{revtex}
\tighten
\begin{document}
\draft
\newcommand{\dd}{{\rm d}}
\newcommand{\D}{{\rm D}}
%\twocolumn[\hsize\textwidth\columnwidth\hsize\csname
%@twocolumnfalse\endcsname
%%%%%%%%%%%%%%%%%%%%%%%%%%%%%%%%%%%%%%%%%%%%%%%%%%%%%%%%%%%%%%%%%%%
\title{Cosmic Strings Lens Phenomenology: General Properties of
Distortion Fields}
\author{Jean--Philippe Uzan$^{1,2}$ and 
        Francis Bernardeau$^3$}
\address{(1) Laboratoire de Physique Th\'eorique, UMR--8627 du CNRS,
             universit\'e Paris XI,\\ b\^atiment 210,
             F--91405 Orsay cedex (France)\\
         (2) D\'epartement de Physique Th\'eorique, Universit\'e 
             de Gen\`eve,\\
             24 quai E. Ansermet, CH--1211 Geneva 4 (Switzerland)\\
         (3) Service de Physique Th\'eorique, CE de Saclay\\
             F--91191 Gif--sur--Yvette Cedex (France).}

\date{\today}
\maketitle
\begin{abstract}
We reconsider the general properties of gravitational lensing effects
induced by cosmic string systems, taking into account their equation of
state and motion equations.  We explicitly show that the deflection
patterns induced by a string is equivalent to the one of a lineic and
static distribution of matter.  We then rigorously show that the
convergence part of the deformation field is always zero, except on the
intersection of the string worldsheet and the observer past light cone,
extending previous results obtained in peculiar cases.  Phenomenological
consequences of this result on multiple image systems are investigated.
\end{abstract}

\pacs{
{\bf Preprint numbers:} LPT--ORSAY 00/33, UGVA--DPT 00/03--1074,
SPhT--Saclay 00/051}

\vskip2pc
%%%%%%%%%%%%%%%%%%%%%%%%%%%%%%%%%%%%%%%%%%%%%%%%%%%%%%%%%%%%%%%%%%%
\section{Introduction}

It  is well  known  that topological  defects  may appear  whenever, in  the
thermal history of the universe,  a symmetry breaking phase transition occurs
\cite{kibble76,mermin},   as  for   instance   in  grand--unified   theories
\cite{kibble76}  or in  some extensions  of the  standard  electro-weak model
\cite{peter92}.   Such  defects  represent  spacetime  positions  where  the
underlying order parameter cannot relax, because of topological constraints,
to its low energy  vacuum state \cite{kibble76,vilenkin}.  They are expected
to interact mainly gravitationally with  the ordinary matter so that they can
induce (i) deflection and  redshifting of massless particles, (ii) accretion
of massive  non--relativistic particles and (iii)  emission of gravitational
waves (see e.g.  \cite{vilenkin,hindmarsh,durrer99rev} for a review of these
effects).

One very interesting example, from a high energy physics point of view
as well as from a cosmological point of view, corresponds to the case of
cosmic strings.  In this case, their phenomenological properties are
determined by the energy density per unit length of a string, $U$.  For
instance the deflection angle is of order $4\pi G U$, $G$ being the
Newton constant.  For defects formed at the grand--unification scale
($T_{\rm GUT}\sim 10^{16}\,~{\rm Gev}$), we expect effects of a
magnitude of $GU\sim 10^{-6}$.  This corresponds for instance to the
magnitude of the cosmic microwave background (CMB) anisotropies induced
through the Kaiser-Stebbins effects \cite{kaiser}. Although topological
defects may have clear observational signature on the CMB sky
\cite{kaiser,btest}, observations seem to disfavor such an origin
\cite{turok,battye,durrer99,uzan99}.

Nonetheless, it does not mean that topological defects do not exist.
Their detection would be of dramatic importance both for astronomy and
particle physics \cite{kibble76,vilenkin,durrer99rev,up} since for
instance estimation or bounds on their density will help constraining
the high energy physics theories predicting their existence.  Definitive
predictions for string properties are however difficult to obtain,
because in particular of the complex evolution equations that may depend
on their microscopic structure through their equation of state.  For the
so-called Goto--Nambu strings (where the energy per unit length $U$ and
the tension $T$ of the string are equal), it was shown that the
spacetime around such straight cosmic string was conical
\cite{vilenkin81,vilenkin84,vilenkin86,gott85,hogan84}.  Such a cosmic
string formed at GUT scale would therefore induce image pairs of distant
objects with angular separation of order
$\theta\sim5.2''\times(GU/10^{-6})$. From a pure phenomenological point
of view such a string is expected to produce lines of double images
\cite{hindmarsh2}.  Recognizing the peculiarity of such a system, it was
later extended to straight cosmic string with different equation of
state \cite{hindmarsh90,peter94} and to a string with a lightlike
current pulse\cite{peter94b}. Moreover numerical simulations for
Goto--Nambu string were also performed in the case of long strings
\cite{delaix1} and cosmic loops \cite{delaix2,delaix3}.  More quantitatively the
prospects for a direct detection of relic string via gravitational
lensing, and in particular the expected number of events, was first
discussed by Hindmarsh \cite{hindmarsh2} who estimated the angular
length of string per unit area on the sky out to a redshift $z$ to be of
order
$$
\theta_{\rm loops}\sim 0.1\nu z^2\,{\rm deg}^{-1}\qquad \theta_{\rm
long\,string}\sim 0.1A z^2\,{\rm deg}^{-1}
$$ 
where  $\nu$  and  $A$  are  two  coefficients  of  order  unity  (see  also
\cite{delaix1}).  In conclusion, it  is widely believed that the observation
of  a  cosmic  string  can  be achieved  through  double  image  detections,
although, in  practice, it might  be difficult to  be positive about  such a
detection since  pairs with  the same angular  separation appearing  by pure
coincidence can  be very high (as pointed  by Cowie and Hu  who reported for
such a cosmic string lens candidate \cite{cowie,emhu}).

The aim of this article and its companion \cite{bu2} is to make a systematic
investigation  of the gravitational  lensing effects  by cosmic  strings. We
focus  our analysis  on the  deformation equation  of a  geodesic  bundle in
presence  of cosmic  strings  (\S~\ref{I}). Standard  approximations of  the
gravitational  lens theory  are also  discussed  in this  section.  After  a
description  of  the cosmic  string  dynamics  in  \S~\ref{II}, we  show  in
\S~\ref{III} that the deflecting potential  of a cosmic string is equivalent
to  the one by  a static  distribution of  matter on  the projection  of the
string worldsheet  onto the observer past  light cone. In the  course in this
calculation we show that the deformation field induced by cosmic strings has
a zero convergence (without  any approximation). Examples illustrating these
results are discussed as well as the validity of the thin lens approximation
and  the influence  of the  equation of  state. In  \S~\ref{IV} we
investigate  the  phenomenological  consequences  of  the  zero  convergence
property on multiple  image systems.  In a companion  paper, \cite{bu2}, we
propose  a  phenomenological  model   of  string  energy  distribution  that
gives a more quantitative account of these results.

%%%%%%%%%%%%%%%%%%%%%%%%%%%%%%%%%%%%%%%%%%%%%%%%%%%%%%%%%%%%%%%%%%%
\section{Evolution of a light beam}\label{I}

In lens systems  that are usually considered in  cosmology, such as galaxies
or galaxy clusters,  the metric perturbations correspond to  those of scalar
perturbations.   This  is not  the  case  for  cosmic string  effects  where
relativistic motions, non trivial equation  of state, also induce vector and
tensor  perturbations.   We are  thus  forced  to  consider the  deformation
equations of light  beams in their full generality.   In the geometric optic
approximation,  an electromagnetic  plane wave  propagating in  an arbitrary
spacetime ${\cal M}$  without interaction with matter can  be described by a
null  geodesic  \cite{mtw}.  The goal  of  this  section  is to  review  the
description of  the evolution and distortion  of a bundle  of null geodesics
and  we start  by introducing  the  standard elements  of the  gravitational
lensing theory and then apply them to a perturbed spacetime. We then discuss
the  thin   lens  approximation   and  finish  by   some  comments   on  its
applicability.

\subsection{Basics of gravitational lensing}

We consider a bundle of null geodesics $g$ propagating in a spacetime ${\cal
M}$.  Each geodesic can be described as
\begin{equation}
g:\quad x^\mu(\lambda)=\bar x^\mu(\lambda)+\xi^\mu(\lambda)
\end{equation}
where $\bar  x^\mu(\lambda)$ is the equation  of a geodesic  $g_0$ chosen as
reference  and  $\xi^\mu$  is  a  displacement  vector  labeling  the  other
geodesics with respect to $g_0$. Greek indices run from 0 to 3 and $\lambda$
is an  affine parameter along the  geodesic $g_0$. With  these notations, we
can define the tangent vector to $g_0$ by
\begin{equation}\label{defkmu}
k^\mu\equiv\frac{\dd\bar x^\mu}{\dd\lambda}.
\end{equation}
It is a null vector satisfying the geodesic equation, i.e solution
of
\begin{equation}\label{eqkmu}
k_\mu k^\mu=0,\quad k^\mu\nabla_\mu k^\nu=0,
\end{equation}
where  $\nabla_\mu$ is  the covariant  derivative associated  to  the metric
$g_{\mu\nu}$ the signature of which will be chosen as $(-,+,+,+)$.

Now, we consider such a bundle converging at a point $O\in{\cal M}$ where we
assume  that there is  an observer  with 4--velocity  $u^\mu$ ($u^\mu$  is a
timelike vector, i.e.  such that $u^\mu u_\mu=-1$) and  we choose the affine
parameter $\lambda$  to vanish at $O$  and to increase toward  the past.  We
then  consider  at   $O$  the  basis  $(k^\mu,u^\mu,n^\mu_1,n^\mu_2)$  where
$n^\mu_{1,2}$  are two  spacelike vectors  ($n^\mu_{a}n_{a\mu}=+1$, $a=1,2$)
such that
\begin{equation}\label{ortho}
n^\mu_{1}n_{2\mu}=0,\quad
k_\mu n^\mu_{a}=0\quad\hbox{and}\quad
u_\mu n^\mu_{a}=0
\end{equation}
and $k^\mu$ is the null vector defined in (\ref{defkmu}). Starting from this
basis  at $O$,  we construct  such a  basis at  any point  of the  light ray
worldline $\bar x^\mu$ by parallelly transporting it as
\begin{equation}\label{Tpara}
k^\mu\nabla_\mu X^\nu=0
\end{equation}
for  $X=u,n_1$ and $n_2$.  Note that  since $k^\mu$  satisfies (\ref{eqkmu})
this implies that  (\ref{Tpara}) is in fact the  Fermi--Walker transport and
thus $u$, $n_1$  and $n_2$ remain orthonormal and  satisfy (\ref{ortho}) for
all     $\lambda$.     Since     the    tangent     vector    $k_g^\mu\equiv
k^\mu+\dd\xi^\mu/\dd\lambda$ to  each geodesic $g$  of the bundle is  a null
vector,     we     deduce     from     $k_g^\mu     k_{g\mu}=g_{\mu\nu}(\bar
x^\alpha+\xi^\alpha)k_g^\mu                  k_g^\nu=0$                 that
$2k_\mu\dd\xi^\mu/\dd\lambda+k^\mu             k^\nu\xi^\alpha\partial_\alpha
g_{\mu\nu}=0$ at first order in $\xi$.  It can then be concluded that, using
(\ref{Tpara}), $k_\mu\xi^\mu$ is constant along the geodesic and vanishes at
$O$ so that it can be decomposed as
\begin{equation}\label{decompxi}
\xi^\mu=\xi_0k^\mu+\sum_{a=1,2}\xi_an^\mu_a.
\end{equation}
$\xi_0$   does  not  vanish   in  general,   but  two   such  decompositions
(\ref{decompxi}) with different $\xi_0$  parameterize the same light ray. We
can  for instance impose  that $\xi^\mu$  is spatial  for the  observer with
4-velocity $u^\mu$  (i.e.  $k_\mu  u^\mu=0$) which then  fixes the  value of
$\xi_0$. We also  decompose the coordinates of every event  of ${\cal M}$ in
the neighborhood of $g_0$ as
\begin{equation}\label{base_dec}
x^\mu=\lambda k^\mu+\sum_{a=1,2}x_a n^\mu_a +\tau u^\mu.
\end{equation}

\begin{figure} 
\centerline{
\epsfig{figure=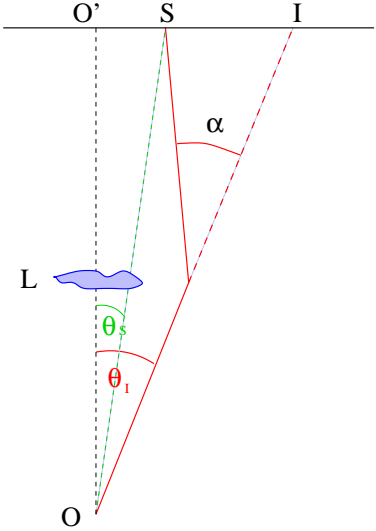,width=5cm}}
\caption{Description of the lensing geometry. $S$ is the source, $L$
the lens and $I$ the image of $S$. $0'S\equiv\xi(\lambda_{\rm S})$.}
\label{alpha} 
\end{figure}

The equation of  evolution of $\xi^\mu$ is obtained  by writing the geodesic
deviation equation \cite{mtw}
\begin{equation}\label{devgeo}
\frac{\D^2}{\dd^2\lambda}\xi^\mu=R^\mu_{\;\nu\alpha\beta}k^\nu
k^\alpha \xi^\beta
\end{equation}
where  $R_{\mu\nu\alpha\beta}$   is  the   Riemann  tensor  of   the  metric
$g_{\mu\nu}$ and where  $\D/\dd\lambda\equiv k^\nu\nabla_\nu$. Inserting the
decomposition  (\ref{decompxi}) in  (\ref{devgeo}) and  using the  fact that
$\xi_a=n_a^\mu\xi_\mu$   is   a   scalar   (so   that   $\D\xi_a/\dd\lambda=
\dd\xi_a/\dd\lambda$ with $\dd/\dd\lambda\equiv k^\mu\partial_\mu$) leads to
\begin{equation}\label{devgeo2}
\ddot\xi_a={\cal R}_a^b\xi_b
\end{equation}
where  ${\cal  R}_{ab}\equiv  R_{\,\mu\nu\alpha\beta}k^\nu k^\alpha  n^\mu_a
n^\beta_b$ and a  dot refers to a derivation with  respect to $\lambda$. Due
to the linearity of the geodesic deviation equation (\ref{devgeo2}), $\xi_a$
can  be  related  to  its  initial value  $\dot\xi_a(0)$  through  a  linear
transformation ${\cal D}_{ab}$ as
\begin{equation}\label{defD}
\xi_a(\lambda)\equiv{\cal D}_{a}^b(\lambda)\dot\xi_b(0).
\end{equation}
Since  $\xi(0)=0$ for  a  bundle  converging at  $O$,  with two  derivatives
(\ref{defD})  and using this  equation again  to eliminate  $\dot\xi(0)$, we
obtain the equation of evolution for ${\cal D}_{ab}$
\begin{equation}\label{evoD}
\ddot{\cal D}_{ab}={\cal R}_a^c{\cal D}_{cb}
\end{equation}
with initial conditions
\begin{equation}
{\cal D}_{ab}(0)=0\quad\hbox{and}\quad \dot{\cal D}_{ab}(0)=I_{ab},
\end{equation}
$I_{ab}$  being the  $2\times2$  identity matrix.   This  equation has  been
derived in e.g. \cite{sachs61,peebles93,schneider,seitz94}.  ${\cal D}_{ab}$
characterizes   the   deformation   field   while   looking   in   different
directions. Quoting  that $\dot\xi(0)\equiv\theta_{\rm I}$  is the vectorial
angle    of   observation   and    $\xi(\lambda_{\rm   S})\equiv\lambda_{\rm
S}\theta_{\rm S}$  where $\theta_{\rm S}$ is the  vectorial angular position
of  the source, equation  (\ref{defD}) can  be rewritten  in terms  of these
angles (see figure~\ref{alpha}) as
\begin{equation}\label{Dtheta}
\theta_{\rm S}^a=\frac{{\cal D}^a_b(\lambda_{\rm S})}{\lambda_{\rm S}}
\theta_{\rm I}^b.
\end{equation}
The amplification matrix ${\cal A}^a_b\equiv{\dd\theta_{\rm
S}^a}/{\dd\theta_{\rm I}^b}$
can be expressed in terms of ${\cal D}$ as
\begin{equation}\label{eq15}
{\cal A}^a_b=\frac{{\cal D}^a_b(\lambda_{\rm S})}{\lambda_{\rm S}}.
\end{equation}
In the following, we decompose it in terms of convergence $\kappa$ and shear
$\vec\gamma\equiv(\gamma_1,\gamma_2)$ as
\begin{equation}\label{dec_A}
{\cal A}_{ab}=\left(\begin{array}{cc}
1-\kappa-\gamma_1&\gamma_2\\
\gamma_2 & 1-\kappa+\gamma_1
\end{array}\right).
\end{equation}

\subsection{Application to a perturbed spacetime}

We now restrict our study to a perturbed spacetime of metric
\begin{equation}
\dd s^2=g_{\mu\nu}\dd x^\mu \dd x^\nu\equiv(\eta_{\mu\nu}+h_{\mu\nu}) 
\dd x^\mu\dd x^\nu,\label{metric} 
\end{equation}
with $\eta_{\mu\nu}$ being the Minkowski metric
$\eta_{\mu\nu}=\hbox{diag}(-,+,+,+)$. We work in harmonic gauge
\begin{equation}
\partial_\nu h^{\mu\nu}=0,\label{gauge}
\end{equation}
so that the Ricci tensor at linear order in the perturbation
$h_{\mu\nu}$ reduces to
\begin{equation}
R_{\mu\nu}=\frac{1}{2}(\partial_t^2-\Delta)h_{\mu\nu},\label{ricci}
\end{equation}
$\Delta$ being the Laplacian $\Delta\equiv\partial_i\partial^i$,
Latin indices running from 1 to 3. The Einstein equations take the
simple form
\begin{equation}\label{einstein}
(\partial_t^2-\Delta)h_{\mu\nu}=16\pi G(T_{\mu\nu}-\frac{1}{2}\eta_{\mu\nu}
T^\lambda_\lambda)\equiv16\pi G{\cal F}_{\mu\nu},
\end{equation}
$T_{\mu\nu}$ being the stress-energy tensor of the matter perturbation.

The solution of this equation can be obtained by means of the Green
functions, ${\cal G}^{(\pm)}$, of the d'Alembertian \cite{schwartz68}
\begin{equation}
\left(\Delta_{\vec x}-\partial_t^2\right)
{\cal G}^{(\pm)}(\vec x',t',\vec x,t)=\delta^{(3)} (\vec
x-\vec x')\delta(t-t') \Longleftrightarrow {\cal G}^{(\pm)}(\vec x',t',\vec
x,t)=-\frac{1}{4\pi}\frac{\delta (t'-t\pm\left|\vec x-\vec x'\right|)}
{|\vec x-\vec x'|},
\end{equation}
so that, using the retarded solution, we
solve the Einstein equations (\ref{ricci}) at
linear order as
\begin{equation}\label{solh}
h_{\mu\nu}(\vec x,t)=4G\int \frac{\dd^3\vec x'}
{|\vec x-\vec x'|}{\cal F}_{\mu\nu}
(\vec x',t-\left|\vec x-\vec x'\right|).
\end{equation}

Now, we can solve (\ref{evoD}) order by order: at zeroth order, ${\cal
R}_{ab}=0$ so that it reduces trivially to
\begin{equation}
{\cal D}_{ab}^{(0)}(\lambda)=\lambda I_{ab};
\end{equation} 
at first order, the equation of evolution (\ref{evoD}) gives
\begin{equation}
\ddot{\cal D}_{ab}^{(1)}(\lambda)=\lambda{\cal R}_{ab}^{(1)}(\lambda)
\end{equation}
from which we deduce that ${\cal D}_{ab}^{(1)}(\lambda)$ is  given by
\begin{eqnarray}
{\cal D}_{ab}^{(1)}(\lambda_{\rm S})&=&\int_0^{\lambda_{\rm S}}
\lambda(\lambda_{\rm S}-\lambda) {\cal R}_{ab}^{(1)}(\lambda)\dd\lambda.
\label{dab1}
\end{eqnarray}
It is interesting to note that while ${\cal D}_{ab}$ is not
symmetric in general, it is symmetric at first order in the perturbation.

Using the expression of the Riemann tensor as
$2R_{\mu\sigma\nu\rho}=h_{\sigma\nu,\mu\rho}+h_{\mu\rho,\sigma\nu}-
h_{\nu\mu,\sigma\rho}-h_{\sigma\rho,\nu\mu}$, we get that
\begin{equation}
{\cal R}^{(1)}_{ab}=\frac{1}{2}h_{\nu\sigma,\mu\rho}k^\nu k^\sigma
n^\mu_a n^\rho_b -\frac{1}{2}\frac{\dd}{\dd\lambda}
\left(\Gamma_{\rho\beta}^\alpha
\eta_{\alpha\mu}k^\beta n^\mu_a n^\rho_b\right)
\end{equation}
where  $k^\mu$, $n_1^\mu$  and $n_2^\mu$  are evaluated  on  the unperturbed
geodesic (and  are thus constant) and  where $\Gamma_{\rho\beta}^\alpha$ are
the Christoffel symbols at first order in the perturbation. Choosing
\begin{equation}
n^\mu_a\equiv\delta^\mu_a
\end{equation}
and defining the deflecting potential $\Phi$ as
\begin{equation}\label{def_phi}
\Phi\left(\vec x,t\right)\equiv
\frac{1}{2}h_{\mu\nu}k^\mu k^\nu,
\end{equation}
where  $t(\lambda)=t_0-x_\parallel(\lambda)$ is the  equation of  the photon
trajectory   [$t_0$   is   the   time   of   the   observation
$t_0=t(\lambda=0)$], we obtain that
\begin{equation}\label{d2ab}
{\cal D}^{(1)}_{ab}(x_a,\lambda_{\rm S})=\left(\int_0^{\lambda_{\rm S}}
\lambda(\lambda_{\rm S}-\lambda)
\partial_{ab}\Phi\left(x_a,x_\parallel(\lambda),t(\lambda)
\right)\dd\lambda\right)+
\int_0^{\lambda_{\rm S}}\lambda(\lambda_{\rm S}-\lambda)\Psi_{ab}\left(x_a,
x_\parallel(\lambda),t(\lambda)\right)\dd\lambda
\end{equation}
where
\begin{equation}\label{defPsi}
\Psi_{ab}\left(\vec x,t
\right)\equiv -\frac{1}{2}\frac{\dd}{\dd\lambda}\left(\Gamma_{a\beta}^\alpha
\eta_{\alpha b}k^\beta\right)
\end{equation}
and where $\partial_a$ refers to a derivative with respect to the
coordinates $x_a$ as defined in (\ref{base_dec}).

\subsection{The thin lens approximation for static distribution of matter}

In  the  thin  lens  approximation,  one  assumes that  the  effect  of  the
deflecting body  takes place over only  a small fraction of  the light path.
This approximation  is usually considered in cases  of scalar perturbations.
The aim of  this paragraph is to recall its derivation  in the standard case
to see  to which extent it applies  for cosmic strings. We  thus assume that
the  lens  is  localised at $\lambda=\lambda_{\rm  L}$  with  an
extension  small  compared  to   $\lambda_{\rm  L}$  and  that  this  matter
distribution   is  static  so   that  ${\cal   F}_{\mu\nu}k^\mu  k^\nu\equiv
\Sigma(\vec          x_\perp,x_\parallel(          \lambda_{\rm         L}))
\delta(x_\parallel(\lambda)-x_\parallel(\lambda_{\rm  L}))$,  where $\Sigma$
is the  surface energy  density.  It follows  that the  deflecting potential
reduces, after integration over the direction of propagation, to
\begin{equation}
\Phi(\vec x_\perp,x_\parallel)=2G\int\frac{\dd^2\vec x_\perp'}{\sqrt{
|\vec x_\perp-\vec x_\perp'|^2+(x_\parallel-
x_\parallel(\lambda_{\rm L}))^2}}\Sigma(\vec 
x'_\perp,x_\parallel(\lambda_{\rm L}))
\end{equation}
where $\vec  x_\perp\equiv(x_1,x_2)$. Since only  $\partial_{ab}\Phi$ enters
the expression  of ${\cal D}_{ab}^{(1)}$  and since this quantity  varies as
$(x_\parallel-x_\parallel(\lambda_{\rm L}))^{-3}$ as  soon as we are looking
close   to  the   string  [i.e.   when  $|\vec   x_\perp-\vec  x_{\perp,{\rm
L}}|\ll(x_\parallel- x_\parallel(\lambda_{\rm L})) $] and we can approximate
the deflecting potential as
\begin{equation}
\Phi(\vec x)=\widetilde\Phi(\vec
x_\perp)\delta(x_\parallel-x_\parallel(\lambda_{\rm L}))
\end{equation}
with
\begin{eqnarray}
\widetilde\Phi(\vec x_\perp)&\equiv&
\int_0^{\lambda_{\rm S}}
\Phi(\vec x_\perp,x_\parallel(\lambda))\dd\lambda\nonumber\\
&=&2G\int\Sigma
\left(\vec x'_\perp,x_\parallel(\lambda_{\rm L})\right)\left[\ln
\left(x_\parallel(\lambda)-x_\parallel(\lambda_{\rm L}) 
+\sqrt{(\vec x_\perp-\vec x_\perp')^2
+(x_\parallel(\lambda)-x_\parallel(\lambda_{\rm L}))^2}
\right)\right]_0^{\lambda_{\rm S}} \dd^2\vec x'_\perp.
\end{eqnarray}
Now, if  we assume that  the impact parameter  is small compared to  the two
distances lens--object and observer--lens, i.e.
\begin{equation}
|\vec x_\perp-\vec x_{\perp,{\rm L}}|\ll (x_\parallel(\lambda_{\rm S})-
x_\parallel(\lambda_{\rm L}),x_\parallel(\lambda_{\rm L})),
\end{equation}
we deduce that  the deflecting potential integrated along  the line of sight
is given by
\begin{equation}\label{44}
\widetilde\Phi(\vec x_\perp)=-4G\int\ln\left|\vec x_\perp-
\vec x_\perp'\right|
\Sigma(\vec x_\perp',x_\parallel(\lambda_{\rm L}))
\dd^2\vec x_\perp'
%=8G\lambda_{\rm S}^2
%\int\ln\left|\vec \theta_{\rm I}-\vec\theta'\right|
%\Sigma(\vec\theta')\dd^2\vec\theta'
\end{equation}
up to  a constant which  depends only on $x_\parallel(\lambda_{\rm  L})$ and
$x_\parallel(\lambda_{\rm  S})$;  we   forget  this  constant  since  ${\cal
D}_{ab}$  involving  only  derivatives   of  $\widetilde\Phi$  and  is  thus
independent of  its value. The second contribution  of ${\cal D}^{(1)}_{ab}$
involves the computation of the  potential $\Psi_{ab}$ and one can show from
(\ref{defPsi}) that  if we  deal only with  scalar perturbations  (i.e. such
that $h_{00}=2\phi$ and  $h_{ij}=2\psi\delta_{ij}$) then $\Psi_{ab}=0$.  For
the vector and tensor perturbations,  $\Psi_{ab}$ does not vanish but in the
thin  lens approximation  its contribution  corresponds to  a  boundary term
(time dependent but identical for all  light rays joining the source and the
observer) which can thus be dropped.

Then, the amplification matrix, in the thin lens approximation, reduces to
\begin{equation}\label{Aabtl}
{\cal A}_{ab}=I_{ab}-4G\frac{\lambda_{\rm S}-\lambda_{\rm
L}}{\lambda_{\rm S}}\lambda_{\rm L}\partial_a\partial_b\int
\ln\left|\vec x_\perp-\vec x_\perp'\right|\Sigma(\vec x'_\perp)
\dd^2\vec x'_\perp
\end{equation}
which   rewrites,   after   the   change  of   variables   $\vec\theta'=\vec
x'_\perp/\lambda_{\rm S}$, as
\begin{equation}\label{eq41}
{\cal A}_{ab}(\lambda_{\rm S})=I_{ab}-4G\frac{\lambda_{\rm S}-
\lambda_{\rm L}}{\lambda_{\rm S}}\partial_{\theta_{\rm I}^a}
\partial_{\theta_{\rm I}^b}\int\ln\left|\vec \theta_{\rm I}-\vec\theta'\right|
\lambda_{\rm L}\Sigma(\vec\theta')\dd^2\vec\theta'.
\end{equation}

Decomposing          $\Sigma(\vec\theta)$          as          $\lambda_{\rm
L}\Sigma(\vec\theta)=\int\mu(s)\delta\left(\vec\theta-\vec\theta_{\rm
string}(s)\right)\dd s$ where $\vec\theta_{\rm string}$ represents the locus
of the  string on  the plane $x_\parallel=x_\parallel(\lambda_{\rm  L})$, we
get that (\ref{eq41}) reduces to
\begin{equation}\label{article2}
{\cal A}_{ab}(\lambda_{\rm S})=I_{ab}-\partial_{\theta_{\rm I}^a}
\partial_{\theta_{\rm I}^b}\varphi(\vec\theta_{\rm I})\quad
\hbox{with}\quad
\varphi(\vec\theta_{\rm I})\equiv4G\frac{\lambda_{\rm S}-
\lambda_{\rm L}}{\lambda_{\rm S}}\int\ln\left|\vec \theta_{\rm I}-
\vec\theta_{\rm string}(s)\right|\mu(s)\dd s
\end{equation} 
where $\mu(s)$  is the projected total  lineic energy density  of the string
(which mixes the  effect of the lineic energy, the  tension and the currents
along the string if any) and $\varphi$ is the effective projected potential.

When dealing with topological defects,  there are different reasons why such
an approximation may not hold. First  the strings are extended and move with
relativistic speed  so that (i)  they are a  priori not confined in  a plane
$\lambda_{\rm  L}\approx\,$constant  and (ii)  one  cannot  assume that  the
distribution of matter of the lens  is static so that the time dependence in
the line-of-sight integration in (\ref{dab1})  has to be taken into
account. These issues will be addressed in \S~\ref{III}.B after a
description of the general stress--energy tensor of strings (\S~\ref{II}).

\subsection{Comments}

\subsubsection{Gravitational potential and deflecting potential}

As a first  comment, let us stress that in  general the deflecting potential
$\Phi$  is  different  from   the  Newtonian  gravitational  potential.  For
instance,  a general  perturbed  spacetime has  the general  post--Newtonian
metric
\begin{equation}\label{metric38}
\dd s^2=-(1-2\phi)\dd t^2+2A_i\dd x^i\dd t+\left[(1+2\psi)\delta_{ij}+
2\bar E_{ij}\right]\dd x^i\dd x^j
\end{equation}
where  $\phi$ and  $\psi$  are  the Newtonian  potentials,  $A_i$ and  $\bar
E_{ij}$  are  the  vector  (rotational)  and  tensor  (gravitational  waves)
perturbations satisfying
\begin{equation}
\bar E_i^i=\partial_i\bar E^{ij}=\partial_iA^i=0.
\end{equation}
It follows that
\begin{equation}
\Phi=\phi+\psi+A_i\gamma^i+\bar E_{ij}\gamma^i\gamma^j
\end{equation}
where  $\gamma^i$ is the  direction of  observation.  This  includes effects
from the rotation of the deflecting body and of gravitational waves. Indeed,
in the case of pure scalar perturbations, we recover that
$$\Phi=2\phi.$$ In  the case of  scalar perturbations, one can  easily check
that $\Psi_{ab}=0$  but topological defects also generate  vector and tensor
perturbations.  In  the thin lens  approximation, the contribution  of these
two  terms reduces  to a  boundary term  that can  be neglected  but  in the
general case of extended object, one has  to check that it is still the case
for the vector and tensor modes.

\subsubsection{Deflection angle}

In  a   general  spacetime,  the   deflection  is  not   straightforward  to
define.  This  is  for instance  the  case  in  a perturbed  spacetime  with
perturbations  on  all scales  (see  e.g.  \cite{durrer94,futamase89} for  a
discussion and a generalization of this concept). If the matter perturbation
causing  the lensing  is localized  in space  then the  metric perturbations
generally die away and the  spacetime is asymptotically unperturbed. In that
case, one  can compute the deflection  angle simply by  solving the geodesic
equation  (\ref{eqkmu})  at  first  order  in the  perturbations.  For  that
purpose, we decompose the tangent vector to the geodesic $g$ as
\begin{equation}
k^\mu_g=\bar k^\mu+\delta k^\mu.
\end{equation}
$\bar  k^\mu$ is  the tangent  vector of  the light  ray in  the unperturbed
Minkowski spacetime  . Note that $\bar  k^\mu$ is different  from the vector
$k^\mu$  defined in (\ref{defkmu})  which labels  the geodesic  of reference
$g_0$. At first order in the  perturbation, it is in fact possible to choose
the unperturbed geodesic as reference  since the displacement $\xi$ is first
order in the perturbation.

Since  both  $\eta_{\mu\nu}\bar   k^\mu\bar  k^\nu=0$  and  $g_{\mu\nu}k^\mu
k^\nu=0$, we  deduce that $\bar k^\mu\delta  k_\mu=0$ at first  order in the
perturbations. At linear order,  the geodesic equation (\ref{eqkmu}) implies
that
\begin{equation}
\frac{\dd}{\dd\lambda}\delta k^\alpha+\delta\Gamma^\alpha_{\mu\nu}
\bar k^\mu\bar k^\nu=0
\end{equation}
with $\delta\Gamma^\alpha_{\mu\nu}\equiv\frac{1}{2}\eta^{\alpha\beta}(
h_{\beta\mu,\nu}+h_{\beta\nu,\mu}-h_{\mu\nu,\beta})$, so that
\begin{equation}
\left[\delta k^\alpha\right]_{\lambda_{\rm S}}^0=-\eta^{\alpha\nu}
\left[h_{\mu\nu}\bar k^\mu\right]_{\lambda_{\rm S}}^0+\frac{1}{2}
\eta^{\alpha\beta}
\int_{\lambda_{\rm S}}^0 h_{\mu\nu,\beta}\bar k^\mu\bar k^\nu \dd\lambda,
\end{equation}
the integral being performed along the unperturbed geodesic.  Forgetting the
boundary term  (which is a time  dependent term but identical  for all light
rays joining the  source and the observer), we can  extract the variation of
the the photon energy measured by an observer with velocity $u^\mu$
\begin{equation}
\delta E=\int_{\lambda_{\rm S}}^0\partial_t\Phi(\vec
x_\perp,x_\parallel(\lambda),t(\lambda))\dd\lambda
\end{equation}
and the deflection $\vec\alpha$ in the plane perpendicular to the line
of sight
\begin{equation}\label{vecgamma}
\vec\alpha=\vec\nabla_\perp\int_{\lambda_{\rm S}}^0\Phi(\vec
x_\perp,x_\parallel(\lambda),t(\lambda))\dd\lambda.
\end{equation}
The effect on  the photon energy, and thus on its  redshift, is nothing else
but the well known Sachs--Wolfe effect \cite{sw}.

Focusing  on (\ref{vecgamma}), since  $\Phi$ is  evaluated along  the photon
geodesic $t(\lambda)=t_0-x_\parallel(\lambda)$, we deduce that
\begin{equation}
\left[\partial_t+\partial_{x_\parallel}\right]\Phi(\vec
x_\perp,x_\parallel(\lambda),t(\lambda))=0
\end{equation}
along the photon path.  Hence,  rewriting the three dimensional Laplacian as
$\Delta= \partial_{x_\parallel}^2+\Delta_\perp$  where $\Delta_\perp$ is the
two dimensional Laplacian and using the Einstein equations (\ref{einstein}),
we deduce that
\begin{equation}
\vec\nabla_\perp.\vec\alpha
=-8\pi G\int_{\lambda_{\rm S}}^0{\cal F}\left(\vec x_\perp,
x_\parallel(\lambda),t(\lambda)\right)\dd\lambda
=-8\pi G\int_{0}^{t(\lambda_{\rm S})-t_0}
{\cal F}\left(\vec x_\perp,x_\parallel(t),t\right)\dd t
\end{equation}
after  choosing  the  parameterization  $\lambda=t_0-t$ and  where  we  have
introduced  ${\cal F}\equiv{\cal F}_{\mu\nu}  k^\mu k^\nu=  T_{\mu\nu} k^\mu
k^\nu$.  It follows that the 2--divergence of the deflection depends only on
the projection of ${\cal F}_{\mu\nu}$  onto the photon trajectory in between
the source and the observer and thus vanishes on all directions which do not
intersect  the string  worldsheet. Note  that such  a result  holds  for any
relativistic and/or extended lens.

Now, in  the thin  lens approximation, one  can relate $\theta_{\rm  S}$ and
$\theta_{\rm I}$ (see figure~\ref{alpha}) by
\begin{equation}\label{48}
\theta_{\rm S}^a=\theta_{\rm I}^a-\frac{\lambda_{\rm S}-
\lambda_{\rm L}}{\lambda_{\rm S}}\alpha^a.
\end{equation}
The  amplification  matrix  being  given by  ${\cal  A}^a_b=\dd\theta^a_{\rm
S}/\dd\theta^b_{\rm I}$, we obtain that
\begin{equation}\label{Aabthinlens}
{\cal A}_{ab}=I_{ab}+\frac{\lambda_{\rm S}- \lambda_{\rm
L}}{\lambda_{\rm S}}\partial_{\theta^b}\partial_a\int_0^{\lambda_{\rm
S}}\Phi\dd\lambda.
\end{equation}
Since the lens is localised in planes close to $\lambda=\lambda_{\rm
L}$, we have that $\partial_{\theta^a}\approx\lambda_{\rm
L}\partial_{a}$ so that the former expression of the amplification
matrix reduces to (\ref{Aabtl}) once we use the expression (\ref{44})
for the integrated potential. In both approaches, we find that the
deflection angle is given by the usual expression (see
e.g. \cite{mellier00})
$$
\vec\alpha=4G\lambda_{\rm L}\int\frac{\vec\theta^{\rm I}-\vec\theta'}
{|\vec\theta^{\rm I}-\vec\theta'|^2} \Sigma(\theta')\dd^2\vec\theta'
$$  
and  we  emphasize  that  defining  the  amplification  matrix  through  the
deflection  angle  implicitely  assumes  that   we  are  in  the  thin  lens
approximation (as seen for instance on equation (\ref{48})).

\subsubsection{Applicability to cosmology}

It has  probably not escaped  a careful reader  that we have  restricted our
calculations   to   perturbations  around   a   Minkowski  spacetime.    The
justification of such  a choice is that the null  geodesics of two conformal
spacetimes are  identical, so that lensing  effects are the  same.  The only
difference  when considering  an expanding  Friedman--Lema\^{\i}tre universe
will  be the computation  of the  distances, i.e.  of $\lambda_{\rm  S}$ and
$\lambda_{\rm  L}$ in  (\ref{eq41}).  Note  also that  the expansion  of the
universe affects the  dynamics of the topological defects  network (see e.g.
\cite{carter}) but this will not be  relevant on time scales of order of the
impact parameter  which are  small compared to  the dynamical scales  of the
universe.

%%%%%%%%%%%%%%%%%%%%%%%%%%%%%%%%%%%%%%%%%%%%%%%%%%%%%%%%%%%%%%%%%%
\section{Dynamics of cosmic strings}\label{II}

The  determination  of the  amplification  matrix  ${\cal  A}$ requires  the
knowledge of stress--energy tensor of  the cosmic strings.  In this section,
we first  present the definition of  this tensor and derive  the equation of
motion of a string. We then  focus on the particular case of a non--rotating
cosmic loop.

\subsection{Equations of motion of the string}

As shown by Carter \cite{carter96}, there  is an elegant way to describe the
dynamics of  a $(d<n)$--brane embedded  in a $n$--dimensional  spacetime. In
this section, we recall the main steps of this formalism necessary to obtain
the  dynamical equation of  evolution of  the string;  details can  be found
in~\cite{carter96}.  It  requires the introduction of the  induced metric on
the string worldsheet
\begin{equation}\label{fab}
f_{AB}\equiv g_{\mu\nu}x^\mu_{,A} x^\nu_{,B},
\end{equation}
where $A,B...$  refers to coordinates on  the worldsheet and  from which one
can construct the fundamental  tensor $\bar\eta^{\mu\nu}$ and the orthogonal
projector $\perp^{\mu\nu}$ as
\begin{equation}\label{bareta}
\bar\eta^{\mu\nu}\equiv f^{AB}x^\mu_{,A} x^\nu_{,B},\quad
\hbox{and}\quad \perp_\nu^\mu\equiv g^\mu_\nu-\bar\eta^\mu_\nu.
\end{equation}
The   covariant  derivative  $\nabla$   defines  a   tangentially  projected
differentiation operator
\begin{equation}\label{tangdiff}
\bar\nabla_\mu\equiv\bar\eta_\mu^\nu\nabla_\nu.
\end{equation}
The second fundamental tensor is defined by
\begin{equation}
{K_{\mu\nu}}^\rho\equiv
\bar\eta^\sigma_\nu\bar\nabla_\mu\bar\eta^\rho_\sigma,
\end{equation}
and  the condition that  the worldsheet  is integrable,  i.e.  that  all its
elements mesh to form a well defined $d$--surface, is expressed by
\begin{equation}\label{weingarten}
{K_{[\mu\nu]}}^\rho=0,\label{eq_geo}
\end{equation}
which is a {\it geometric identity} for the worldsheet.

In  the  case  of  a string  ($d=2$),  the  Lagrangian  density
$\widehat{\cal L}$ can be expressed as
\begin{equation}
\widehat{\cal L}=\frac{1}{\sqrt{-g}}\int\sqrt{f}{\rm d}^2\zeta\bar{\cal L}
\delta^{(4)}\left[x^\mu-x^\mu(\zeta^A) \right]
\end{equation}
where $f$ is the determinant  of the metric $f_{AB}$ defined in (\ref{fab}).
$\widehat{\cal  L}$  is  distributional  and  not  confined  to  the  string
worldsheet whereas  $\bar{\cal L}$  is locally regular  but confined  on the
string worldsheet. The string action can  then be written either in terms of
$\widehat{\cal L}$ or of $\bar{\cal L}$ as
\begin{equation}\label{action}
S_{\rm string}=\int{\rm d}^2\zeta\sqrt{f}\bar{\cal L}
              =\int{\rm d}^4x\sqrt{-g}\widehat{\cal L}.
\end{equation}
Now,  making an infinitesimal  variation $\delta  g_{\mu\nu}$ of  the action
(\ref{action}) provides a definition of the ``surface'' stress energy tensor
density  $\widetilde T^{\mu\nu}$ (confined  and regular)  and of  the stress
energy tensor density  $\widehat T^{\mu\nu}$ (distributional and unconfined)
by
$$
2\delta S_{\rm string}=
\int{\rm d}^2\zeta\sqrt{f}\widetilde T^{\mu\nu}\delta g_{\mu\nu}
=\int{\rm d}^4x\sqrt{-g}\widehat T^{\mu\nu}\delta g_{\mu\nu}.
$$
These two stress--energy tensor are related by
\begin{equation}\label{tmunu2}
\widehat T^{\mu\nu}=\frac{1}{\sqrt{-g}}\int{\rm d}^2\zeta\sqrt{f}\widetilde T^{\mu\nu}
\delta^{(4)}\left[x^\mu-x^\mu(\zeta^A) \right].
\end{equation}
The internal coordinate stress  energy tensor, $\widetilde T^{AB}$, has been
projected onto a corresponding  background stress energy tensor, $\widetilde
T^{\mu\nu}$, as in  (\ref{bareta}). One can then show  that the general form
of the {\it dynamical equation} of the string is
\begin{equation}\label{eq_dyn0}
\bar\nabla_\mu\widetilde T^{\mu\nu}=f^\nu
\end{equation}
$f^\nu$ being the  force exerted on the string by  any background field such
as e.g. an electromagnetic field. This dynamical equation (\ref{eq_dyn0}) is
equivalent to the more  natural equation of conservation $\nabla_\mu\widehat
T^{\mu\nu}=\hat  f^\nu$   with  $\hat  f^\nu$  related  to   $f^\nu$  as  in
(\ref{tmunu2}).

For a string of energy per unit length $U$ and of tension $T$, the surface
stress energy tensor density is of the form
\begin{equation}\label{tmunu1}
\widetilde T^{\mu\nu}=U u^\mu u^\nu - T v^\mu v^\nu,
\end{equation}
where $u^\mu$ and $v^\mu$ are respectively a timelike ($u^\mu u_\mu=-1$) and
a spacelike ($v^\mu v_\mu=+1$) unit  vector tangent to the string worldsheet
(i.e.    $\perp^\mu_\nu    u^\nu=    \perp^\mu_\nu   v^\nu=0$)    so    that
$\bar\eta^{\mu\nu}= -  u^\mu u^\nu +  v^\mu v^\nu$.  The  dynamical equation
governing the evolution of the  string is given by the tangential projection
of (\ref{eq_dyn0}) which, in the free case we are considering, leads to
\begin{equation}\label{eq_dyn}
\eta^\rho_\nu\bar\nabla_\mu\widetilde T^{\mu\nu}=0.
\end{equation}
This equation  of evolution  (\ref{eq_dyn}) can then  be solved once  we are
given an equation  of state, i.e.  $U(T)$. Such  equations are provided once
the microscopic  structure of the string  is described. The  most well known
are the  Goto--Nambu strings  \cite{goto71} ($U=T$) and  the Nielsen--Olesen
strings  \cite{nielsen}  ($U+T=const.$)  and  some have  been  obtained  for
superconducting strings \cite{supercond}.

\subsection{Application to a non rotating cosmic string loop}

In  the case  of a  non rotating  circular loop  of radius  $R$, we  work in
cylindrical coordinates $(t,r,\theta,z)$ and assume  that it is lying in the
plane $z=z_s$. Parametrising the loop worldsheet as
\begin{equation}\label{param}
t_{\rm loop}=t, \quad\vec r_{\rm loop}\equiv \vec r_\perp=
R(t)(\cos\theta,\sin\theta),\quad x_3\equiv z_{\rm loop}-z_s=0,
\end{equation}
one can show that the unit spacelike vector tangent to the string
worldsheet is $v^\mu=\theta^\mu$ and we have
\begin{equation}\label{64}
u_\mu=\gamma\delta_\mu^t+\gamma\dot R\delta_\mu^r,\quad
\theta_\mu=R\delta_\mu^\theta,
\end{equation}
with $\gamma\equiv  (1-\dot R^2)^{-1/2}$ being the  Lorenz factor associated
with  the   radial  contraction/expansion  of  the   string.   They  satisfy
$\perp^\mu_\nu  u^\nu=\perp^\mu_\nu \theta^\nu=0$.  From  (\ref{tmunu1}) and
(\ref{tmunu2}),  the stress--energy tensor  entering the  Einstein equations
(\ref{einstein}) is given by
\begin{equation}\label{65}
T^{\mu\nu}(\vec x,t)=\left(
\begin{array}{cccc}
\gamma^2U        &\gamma^2\dot R U   &  0   &0 \\
\gamma^2\dot R U &\gamma^2\dot R^2 U &   0  & 0\\
          0       &     0              &-R^2T&0 \\
          0       &          0         &   0  & 0
\end{array}\right)\delta(z-z_s)\delta(r-R).
\end{equation}
We now need to write  the dynamical evolution equation (\ref{eq_dyn}) to get
the evolution of the  radius of the loop as a function  of time.  Using that
$\bar\eta_{\mu\nu}=-u_\mu  u_\nu+\theta_\mu\theta_\nu$  and  the  expression
(\ref{tmunu1})  of the  stress--energy tensor  of the  string,  the equation
(\ref{eq_dyn}) can be rewritten as
\begin{equation}
(-u_\mu u^\nu+\theta_\mu\theta^\nu)\nabla_\nu\left(
U u^\mu u^\rho -T\theta^\mu\theta^\rho \right)=0,
\end{equation}
which reduces to
\begin{eqnarray}
\frac{\dd}{\dd t}(\gamma UR)&=&0\label{eq1}\\
\frac{\dd^2}{\dd t^2}R&=&-\frac{1}{\gamma^2R}\frac{T}{U}.
\label{eq2}
\end{eqnarray}
This system of equations for ($U,T,R$)  is not closed and can be solved when
one specifies an  equation of state $U(T)$. Equation  (\ref{eq1}) shows that
the total energy $\gamma RU$ of the loop is a constant of motion.  Note that
we  have  not used  the  identity  (\ref{weingarten})  which is  identically
satisfied in our present example.

%%%%%%%%%%%%%%%%%%%%%%%%%%%%%%%%%%%%%%%%%%%%%%%%%%%%%%%%%%%%%%%%%%%%%%
\section{Lensing by a cosmic string}\label{III}

\subsection{A first example}\label{first}

As  a first application,  let us  consider a  static straight  cosmic string
lying along  the axis $x_2$  in a plane  perpendicular to the line  of sight
(direction $x_3$ on figure~\ref{tilt} with $\varphi=0$) so that
\begin{equation}\label{cs}
{\cal F}_{\mu\nu}=\frac{1}{2}\left(\begin{array}{cccc}
U-T&&&\\ &U+T&&\\ &&U-T&\\ &&&U+T\end{array}\right)
\delta(\lambda_{\rm L}\theta_1)
\delta(x_\parallel-x_\parallel(\lambda_{\rm L})),
\end{equation}
with  $U$ and $T$  depending on  $x_2=\lambda_{\rm L}\theta_2$  only.  Since
${\cal        F}=U(x_2)\delta(x_1)\delta(x_\parallel-x_\parallel(\lambda_{\rm
L}))$, the  first integral of  (\ref{d2ab}), after integration  over $x'_1$,
reduces to
$$
{\cal A}_{ab}=I_{ab}+2\frac{G}{\lambda_{\rm S}}\int U(x'_2)\dd x'_2
\partial_{ab} J((x_2-x'_2)^2+x_1^2)
$$
with
$$
J((x_2-x'_2)^2+x_1^2)=\int_0^{\lambda_{\rm      S}}\frac{\lambda(\lambda_{\rm
S}-\lambda)}{\sqrt{(x_2-x'_2)^2+x_1^2+(\lambda-\lambda_{\rm           L})^2}}
\dd\lambda
$$    
where     we     have     chosen     a     parametrisation     such     that
$x_\parallel(\lambda)=\lambda$.   This latter integral  can be  computed and
gives
\begin{eqnarray}
J(A)&=&\frac{1}{2}(\lambda_{\rm S}-3\lambda_{\rm L})
\sqrt{A+(\lambda_{\rm S}-\lambda_{\rm L})^2}-
\left(\lambda_{\rm S}-\frac{3}{2}\lambda_{\rm L}\right)
\sqrt{A+\lambda_{\rm L}^2}
\nonumber\\
&&\qquad\qquad+(\lambda_{\rm S}\lambda_{\rm L}-\lambda_{\rm L}^2-\frac{1}{2}A)
\ln\frac{\lambda_{\rm S}-\lambda_{\rm L}+\sqrt{A+(\lambda_{\rm
S}-\lambda_{\rm L})^2}}{-\lambda_{\rm L}+\sqrt{A+\lambda_{\rm L}^2}}
\nonumber
\end{eqnarray}
with $A\equiv(x_2-x'_2)^2+x_1^2$.  It follows that  the amplification matrix
is given by
\begin{equation}\label{67}
{\cal A}_{ab}=I_{ab}+2\frac{G}{\lambda_{\rm S}}\int U(x'_2)\dd x'_2
\left(2J'(A)\delta_{ab}+(x_a-x'_a)(x_b-x'_b)J''(A)\right)
\end{equation}
where $J'\equiv\dd J/\dd A$ and $x'_1=0$. Now, one can estimate the dominant
term in the  integral of expression (\ref{67}) when we  are looking close to
the    string    (i.e.    when    $x_1,x_2\ll\lambda_{\rm    S}-\lambda_{\rm
L},\lambda_{\rm  L}$).  For  that   purpose  we  assume  that  $\lambda_{\rm
S}-\lambda_{\rm  L}\sim\lambda_{\rm L}$  and  set $\lambda\simeq\lambda_{\rm
S}-\lambda_{\rm  L}\simeq\lambda_{\rm L}$.   We then  split the  integral of
expression     (\ref{67})     in     a     contribution     where     $A\sim
x_2^{\prime2}<\lambda^2$        and        another       where        $A\sim
x_2^{\prime2}>\lambda^2$. Using the expansion of $J(A)$ in these two regimes
as
$$
J(A)=-\left(\lambda_{\rm S}-\lambda_{\rm L}\right)\left\lbrace
\begin{array}{ll}
\lambda_{\rm L}\ln(A)+constant+{\cal O}\left(\frac{A}{\lambda^2}\right)
&A/\lambda^2<1\\
\sqrt{A}\left(1+{\cal O}\left(\frac{\lambda}{\sqrt{A}}\right)\right)
&A/\lambda^2>1
\end{array}\right.
$$
one  can show  that, as  long as  we are  looking close  to the  string, the
dominant contribution to the amplification matrix is given by
\begin{equation}
{\cal A}_{ab}\simeq I_{ab}-4G\frac{\lambda_{\rm S}-\lambda_{\rm
L}}{\lambda_{\rm S}}\partial_{\theta^a}\partial_{\theta^b}
\int\ln\left|\vec\theta^I-\vec\theta'\right|
U(\theta'_2)\dd\theta'_2.
\end{equation}
In  the particular case  where $U$  is constant,  this approximate  give the
general   result   of  the   deflection   by   a   straight  cosmic   string
\cite{hindmarsh2,hindmarsh90,peter94} and one can  thus thing that this is a
good approximation  when $U$ is fluctuating  around a mean  value. Note also
that for such an infinite string  lying in a plane perpendicular to the line
of sight  we recover  the general form  (\ref{article2}) of  the deformation
matrix in the thin lens approximation.

In  the following  of  this  article, we  investigate  more general  results
concerning the deformation fiels by a cosmic string which do not assume that
we are in the thin lens regime.

\subsection{General Results}

In the  general case, the source term  generated by a cosmic  string will be
localized on the string worldsheet so that [see equation (\ref{tmunu1})]
\begin{equation}
{\cal F}(\vec x,t)=\int \dd\zeta\widetilde{\cal F}(\vec x,t)\delta
\left(\vec x-\vec r(\zeta,t)\right)
\end{equation}
where $(t,\vec r(\zeta,t))$ is  a parameterization of the string worldsheet;
$\vec r(\zeta,t)$ represents  the locus of the string  on each constant time
hypersurface and $\widetilde{\cal F}$ is  the energy density per unit length
[note  that we  have chosen  a  parametrisation such  that $t$  is both  the
coordinate time and  an intrinsic coordinate of the  string worldsheet which
implies that we  have a one dimensional integration  on the spatial internal
coordinate   $\zeta$  and   not  a   two  dimensional   integration   as  in
(\ref{tmunu1})].  Inserting this  decomposition in  (\ref{solh}),  we deduce
that the  deflecting potential  (\ref{def_phi}) is given,  after integration
over space, by
\begin{equation}
\Phi(\vec x,t)=2G\int\frac{\dd t'\dd\zeta}{\left|
\vec x-\vec r(\zeta,t')\right|}
\widetilde{\cal F}\left[\vec r(\zeta,t'),t'\right]
\delta\left(t'-t+\left|\vec x-\vec r(\zeta,t')\right|\right).
\end{equation}
Following \cite{stebbins88},  the integration over $t'$ can  be performed by
introducing $t_{\rm string}(\vec x,\zeta,t)$ solution of
\begin{equation}
t-t_{\rm string}(\vec x,\zeta,t)=\left|\vec x-
\vec r\left(\zeta,t_{\rm string}(\vec x,\zeta,t)\right)\right|
\end{equation}
so that
$$
\delta\left(t'-t+\left|\vec x-\vec r(\zeta,t')\right|\right)=
\frac{\left|\vec x-\vec r(\zeta,t')\right|}{\left|\vec x-\vec
r(\zeta,t')\right|-\partial_t{\vec r}(\zeta,t').(\vec x-
\vec r(\zeta,t'))}\delta\left(t'-t_{\rm string}\right)
$$
from which we deduce that
\begin{equation}\label{76}
\Phi(\vec x,t)=2G\int\frac{\widetilde{\cal F}\left[\vec r(\zeta,t_{\rm
string}(\vec x,\zeta,t)),t_{\rm string}(\vec
x,\zeta,t)\right]}{\left|\vec x-\vec r(\zeta,t_{\rm string}(\vec
x,\zeta,t))\right|-\partial_t{\vec r}(\zeta,t_{\rm string}(\vec
x,\zeta,t)).(\vec x- \vec r(\zeta,t_{\rm string}(\vec x,\zeta,t)))}\dd\zeta.
\end{equation}
$\Phi$ on  the point  $(\vec x,t)$ is  then given  by the projection  of the
string  energy on  the past  light cone  of this  point, i.e.  on  the curve
$\left\lbrace\vec     r(\zeta,t_{\rm     string}(\vec     x,\zeta,t)),t_{\rm
string}(\vec  x,\zeta,t)\right\rbrace$  which  is  the intersection  of  the
string worldsheet  with the past light  cone of the event  $(\vec x,t)$ [see
figure~\ref{intersection}].

Now, focusing  on $\kappa$,  the deformation matrix  is explicitly  given by
(\ref{def_phi})  and  (\ref{d2ab}) and,  proceeding  as  for the  deflection
angle, one can easily sort out that
\begin{eqnarray}
\delta^{ab}{\cal D}^{(1)}_{ab}&=&\int_0^{\lambda_{\rm S}}
\lambda(\lambda_{\rm S}-\lambda)\Delta_\perp\Phi(\vec
x_\perp,x_\parallel(\lambda),t(\lambda))\dd\lambda
+\int_0^{\lambda_{\rm S}} \lambda(\lambda_{\rm S}-\lambda)\delta^{ab}\Psi_{ab}
\dd\lambda
\nonumber\\
&=&\int_0^{\lambda_{\rm S}}\lambda(\lambda_{\rm S}-\lambda)
\left[\left(\partial_t^2-\partial_\parallel^2\right)\Phi(\vec
x_\perp,x_\parallel(\lambda),t(\lambda))-8\pi G{\cal F}(\vec
x_\perp,x_\parallel(\lambda),t(\lambda))\right]\dd\lambda\nonumber\\
&+&\int_0^{\lambda_{\rm S}} \lambda(\lambda_{\rm S}-\lambda)\delta^{ab}\Psi_{ab}
\dd\lambda\label{71}
\end{eqnarray}
The first term  of the first integral vanishes when  evaluated on the photon
geodesic so that  the contribution of the first  integral to the convergence
$\kappa$ defined in (\ref{dec_A}) reduces to
\begin{eqnarray}\label{kappa}
\kappa(\vec x_\perp,t_0)=4\pi G\int_0^{\lambda_{\rm S}}\frac{\lambda
(\lambda_{\rm S}-\lambda)}{\lambda_{\rm S}}
{\cal F}(\vec x_\perp,x_\parallel(\lambda),t(\lambda))\dd\lambda
=4\pi G\int_{t_{\rm emission}}^{t_0}
\frac{(t-t_{\rm emission})(t_0-t)}{t_0-t_{\rm emission}}
{\cal F}(\vec x_\perp,x_\parallel(t),t)\dd t.
\end{eqnarray}
where we recall  that $t_0\equiv t(\lambda=0)$ is the  time of reception and
where $t_{\rm  emission}\equiv t(\lambda_{\rm S})$ is the  time of emission.
The contribution  of the  second integral of  (\ref{71}) vanishes  since (i)
$\Psi_{ab}\delta^{ab}=0$ both for scalar perturbations [$\phi$ and $\psi$ in
(\ref{metric38})] and  for vector perturbations  [$A_i$ in (\ref{metric38})]
and   (ii)   for   tensor   modes  [$\bar   E_{ij}$   in   (\ref{metric38})]
$\Psi_{ab}\delta^{ab}\propto\frac{\dd}{\dd\lambda}\left[(\partial_t+
\partial_\parallel)(\bar E^1_1+\bar  E^2_2)\right]=0$ when evaluated  on the
photon trajectory.   We conclude that  the convergence $\kappa$ is  given by
(\ref{kappa}). It is  then given by the distribution  of matter evaluated on
the photon trajectory, up to a  geometrical factor.  The lensing by a cosmic
string  is  thus equivalent  to  the lensing  by  a  linear distribution  of
matter. {\em As a consequence, $\kappa=0$ everywhere} but on directions such
that the  observer past  light cone intersects  the string  worldsheet; this
result, valid whatever the equation of  state, is one of the main results of
this article.  It holds for any relativistic  lens and does not  rely on the
thin lens approximation.

For instance if the string is  lying in a plane perpendicular to the line of
sight then (\ref{kappa}) reduces to
\begin{eqnarray}\label{kappa2}
\kappa(\vec x_\perp,t_0)=4\pi G\frac{\lambda_{\rm S}-\lambda_{\rm L}}{
\lambda_{\rm S}}\lambda_{\rm L}\Sigma
(\vec x_\perp,x_\parallel(\lambda_{\rm L}),t(\lambda_{\rm L}))
\propto \nabla_\perp.\vec\alpha.
\end{eqnarray}
Under this form,  again we see that $\kappa=0$  everywhere but on directions
intersecting the string worldsheet.

Note that the drivation of (\ref{kappa}) relies strongly on the fact
that we took the trace of the deformation matrix in (\ref{71}). Therefore,
similar expression to (\ref{kappa}) cannot be obtained for the other
components of the amplification matrix and one has to rely on (\ref{76}).

It has also to be noted that the expression of the amplification matrix
in (\ref{Aabthinlens}) (which relies on the thin lens approximation)
together with the general result ({\ref{kappa}) implies that the
phenomenological description of ${\cal A}^a_b$ in (\ref{article2}) is
very general.  It holds for any string dynamics provided the extension
of the string is small enough for the thin lens approximation to hold.
Different aspects of these results are illustrated in the next
paragraphs. A more elaborate phenomenological investigation based on
Eq. (\ref{article2}) is proposed in a companion paper \cite{bu2}.

\begin{figure} 
\centerline{
\epsfig{figure=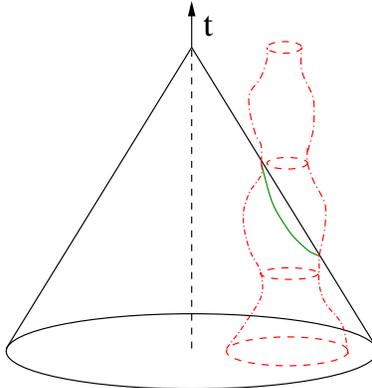,width=5cm}}
\caption{The intersection of the loop worldsheet and the past light cone
of the event ($\vec x,t)$. The dash circles represent the loop in
different constant time hypersurfaces and the dash--dot lines
 the loop worldsheet.}
\label{intersection} 
\end{figure}

\subsection{Lensing by a non--rotating cosmic string loop perpendicular
to the line of sight}

We now consider as an application the case of a non--rotating circular
loop oscillating in a plane perpendicular to the line of sight. Its
dynamics is described by the set of equations (\ref{eq1}--\ref{eq2}) and
from its parameterization (\ref{param}) we deduce that
\begin{equation}
{\cal F}\left(\vec x,t\right)=\int_0^{2\pi} R(t) \gamma(t)U(t)
\delta(x_3)\delta(\vec x_\perp-\vec r_\perp)\dd\theta.
\end{equation}
The deflecting potential (\ref{def_phi}) integrated along the line of sight is then
given by
\begin{eqnarray}\label{iphi}
I[\Phi]&\equiv&\int_{0}^{\lambda_{\rm S}}\Phi\left(\vec x(\lambda),
t(\lambda)\right)\dd\lambda\nonumber\\
&=&2G\int_{-x_{\rm L}}^{x_{\rm S}-x_{\rm L}} \dd x_3\int_0^{2\pi}
\dd\theta\int_{R^2} \dd^2\vec x'_\perp
\int_{-\infty}^{+\infty} \dd t' \gamma RU\frac{\delta\left(t'+x_3-\tau_0+\sqrt{\left(
\vec x_\perp-\vec x'_\perp\right)^2+x_3^2}  \right)}{\sqrt{\left(\vec x_\perp-
\vec x'_\perp\right)^2+x_3^2}}\delta\left(\vec x'_\perp-\vec r_\perp(t',\theta)\right)
\end{eqnarray}
where we have parameterized the geodesic as $t=\tau_0-x_3$ with $\tau_0\equiv t_0-x_{\rm L}$.
Now, defining the new variable $z$ as
\begin{equation}
z\equiv x_3+\sqrt{\left(\vec x_\perp-\vec x'_\perp\right)^2+x_3^2}
\end{equation}
which satisfies
\begin{equation}
\frac{\dd z}{z}=\frac{\dd x_3}{\sqrt{
\left(\vec x_\perp-\vec x'_\perp\right)^2+x_3^2}}.
\end{equation}
The integrated potential (\ref{iphi}) reduces  to
\begin{equation}
I[\Phi]=2G\int_0^{2\pi}\dd\theta\int_{R^2} \dd^2\vec x'_\perp\int_A^B
\frac{\dd z}{z}
\int_{-\infty}^{+\infty} dt' \gamma RU \delta(t'+z-\tau_0)
\delta\left(\vec x'_\perp-\vec r_\perp(t',\theta)\right)
\end{equation} 
where the limits of integration are 
\begin{eqnarray}
A&=&-x_{\rm L}+\sqrt{\left(\vec x_\perp-\vec x'_\perp\right)^2+x_{\rm L}^2}
\simeq\frac{1}{2}\frac{\left(\vec x_\perp-\vec x'_\perp\right)^2}{x_{\rm L}}\\
B&=&x_{\rm S}-x_{\rm L}+\sqrt{\left(\vec x_\perp-\vec x'_\perp\right)^2+
(x_{\rm S}-x_{\rm L})^2}\simeq 2(x_{\rm S}-x_{\rm L}).
\end{eqnarray}
The approximate values  of $A$ and $B$ are obtained at  lowest order when we
consider  zones  close  to  the  string  in  comparison  with  the  distance
string--observer and string--source.
 
After  integration  over  $t'$  and   using  the  loop  equation  of  motion
(\ref{eq1}--\ref{eq2}) which state that $\gamma RU$ is constant, we get that
\begin{equation}\label{facto}
I[\Phi]=2G\gamma RU J(\vec x_\perp,t_0),
\end{equation}
where $J(\vec x_\perp,t_0)$ is a dimensionless geometrical integral given by
\begin{equation}\label{f2}
J(\vec x_\perp,t_0)=\int_0^{2\pi}\dd\theta\int_{R^2} \dd^2\vec x'_\perp
\int_{\frac{1}{2}\frac{\left(\vec x_\perp-\vec x'_\perp\right)^2}{x_{\rm L}}}^{2
(x_{\rm S}-x_{\rm L})}\frac{\dd z}{z}\delta\left(\vec x'_\perp-\vec
r_\perp(-z-\tau_0,\theta)\right).
\end{equation}
This integral can be rewritten as
\begin{equation}\label{fu}
J(\vec x_\perp,t_0)=\int_0^{2(x_{\rm S}-x_{\rm L})}\frac{\dd z}{z}
\int_0^{2\pi}\dd\theta\int_{\left|\vec x_\perp-\vec 
x'_\perp\right|^2<2x_{\rm L}z} 
\dd^2\vec x'_\perp \delta\left(\vec x'_\perp-
\vec r_\perp(-z-\tau_0,\theta)\right).
\end{equation}
We deduce that, at $z$ constant, the integral over $x_\perp'$ reduces to the
computation  of  the angle  $\beta$  (see  figure~\ref{triangle}) of  string
within  the disk of  radius $\sqrt{2x_{\rm  L}z}$ which  can be  computed as
followed:

Setting $u\equiv\sqrt{2x_{\rm L}z}$ and $h\equiv OH$, we deduce from
\begin{equation}
u=u_1+u_2,\quad u_1^2+h^2=\vec x_\perp^2,\quad
u_2^2+h^2=R^2
\end{equation}
where $u_1\equiv CH$ and $u_2\equiv HD$ on figure~\ref{triangle} and
$R$ now stands for $R(t=\tau_0)$, that
\begin{equation}
u_1=\frac{\vec x_\perp^2-R^2+u^2}{2u}\quad
u_2=\frac{-\vec x_\perp^2+R^2+u^2}{2u}\quad
\end{equation}
and thus that
\begin{equation}\label{al}
\cos\beta=\frac{\vec x_\perp^2+R^2-u^2}{2|\vec x_\perp|R}.
\end{equation}
The  integral  over  $\theta$  and  $\vec  x_\perp'$  obviously
vanishes  if  $u<\left||\vec  x_\perp|-R\right|$,  reduces  to  $2\pi$  when
$u>|\vec x_\perp|+R$ and gives  $2\beta(u)$ otherwise. Then, after splitting
the  integral  over  $u$  in  (\ref{fu}) in  three  pieces  ($[0,\left||\vec
x_\perp|-R\right|]$, $[\left||\vec x_\perp|-R\right|, |\vec x_\perp|+R]$ and
$[|\vec x_\perp|+R,2\sqrt{x_{\rm L}(x_{\rm S}-x_{\rm L})}]$) we get
\begin{eqnarray}\label{99}
J(\vec x_\perp,t_0)&=& \int_{\left||\vec x_\perp|-R\right|}^{|\vec x_\perp|+R}
2\beta(u)\frac{\dd u}{u}+
2\pi \int_{|\vec x_\perp|+R}^{2\sqrt{x_{\rm L}(x_{\rm S}-x_{\rm L})}}
\frac{\dd u}{u},
\end{eqnarray}
with $\beta$ given  by (\ref{al}).  We can compute this  integral in the two
following regimes
\begin{enumerate}
\item If $|\vec x_\perp| < R(\tau_0)$, (\ref{99}) can be rewritten as
\begin{equation}\label{approxj}
J=2|\vec x_\perp|R\int_{-1}^1\frac{\hbox{Arccos}v}{|\vec x_\perp|^2+R^2
-2|\vec x_\perp|Rv}\dd v+2\pi\ln\left[
\frac{2\sqrt{x_{\rm L}(x_{\rm S}-x_{\rm L})}}{|\vec x_\perp|+R}\right]
\end{equation}
which can be computed to give
\begin{equation}\label{J90}
J = C_1
\end{equation}
where  $C_1$  is  a constant  depending  on  $x_{\rm  L}$, $x_{\rm  S}$  and
$R(\tau_0)$. Then,  there is no deflection  of a light ray  passing inside a
large loop, as first pointed out  in \cite{hogan84} and as expected from the
Gauss theorem.
\item If $|\vec x_\perp| > R(\tau_0)$, (\ref{approxj}) now gives after
integration
\begin{equation}\label{J91}
J= C_1-2\pi\ln\frac{|\vec x_\perp|}{R}
\end{equation}
and we conclude that a small loop perpendicular to the line of sight acts as
a  point mass $M=2\pi\gamma  RU$ {\it  whatever} its  equation of  state. We
checked that this is also valid for a tilted circular loop and it is natural
to expect  that the  fact that a  loop acts  as a point  mass at  a distance
larger than  its caracteristic  size is valid  whatever the geometry  of the
loop.

One  can also  check from  (\ref{J90}--\ref{J91}) that  $\Delta_\perp J(\vec
x_\perp,t_0)=0$  if   $\vec  x_\perp\not=R(\tau_0)$.   Since  $\kappa\propto
\Delta_\perp  J$,  we  recover   the  result  from  (\ref{kappa})  that  the
convergence vanishes if $\vec x_\perp\not=\vec r_\perp$.
\end{enumerate}

\begin{figure} 
\centerline{
\epsfig{figure=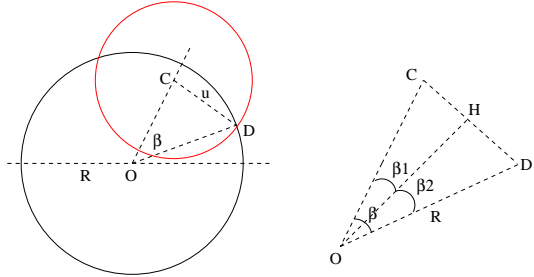,width=7cm}}
\caption{Integration over the disk of radius $u\equiv\sqrt{2x_{\rm L}z}$
around the point $C\equiv\vec x_\perp$.}
\label{triangle} 
\end{figure}

\subsection{Tilted static straight cosmic string}

To finish,  let us consider a  tilted static straight  cosmic string aligned
along  an axis  making  an angle  $\varphi$  with the  direction $x_2$  [see
figure~\ref{tilt}] and for which, from (\ref{cs}),
\begin{equation}
{\cal F}=\int\left[U(\ell)-T(\ell)\sin^2\varphi\right]\dd\ell
\delta(\vec x-\vec r(\ell))
\end{equation}
where $\vec r(\ell)$ is a parameterization of the string. If we choose
$\ell$ such that
\begin{equation}
\vec r(\ell):\quad r_1=0,\quad r_2=\ell\cos\varphi,\quad
r_3=x_\parallel(\lambda_{\rm L})-\ell\sin\varphi,
\end{equation}
the deflecting potential (\ref{76}) is given by
\begin{equation}
\Phi(\vec x)=2G\int
\frac{\dd\ell}{\left|\vec x-\vec r(\ell)\right|}
\left[U-T\sin^2\varphi\right](\ell).
\end{equation}
When $U$ and $T$ are constant, it can be integrated to get
\begin{equation}
\Phi(\vec x)=2G\left[U-T\sin^2\varphi\right]\left(C_\infty-
\ln\left[\vec x^2-(x_2\cos\varphi-x_3\sin\varphi)^2
\right]\right)
\end{equation}
where  $C_\infty$ is  an  infinite  constant and  where  we have  introduced
$x_3\equiv  x_\parallel-x_{\rm L}$.   The infinite  constant  $C_\infty$ can
then  be  forgotten  because  only  the derivatives  of  $\Phi$  enters  the
computation of  the deflection angle  and of the amplification  matrix which
are the observable quantities.  The deflection angle is then given by
\begin{equation}
\vec\alpha=-\nabla_\perp\int_{-x_{\rm L}}^{x_{\rm S}-x_{\rm L}}
\Phi(x_1,x_2,x_3) \dd x_3.
\end{equation}
After integration over $x_3$, we get
\begin{eqnarray}
\alpha_1&=&\frac{4 G\left[U-T\sin^2\varphi\right]}{\cos\varphi}
\left[{\rm arctan}\left(\frac{x_2\sin\varphi+x_3\cos\varphi}{x_1}\right)
\right]_{x_3=-x_{\rm L}}^{x_3=x_{\rm S}-x_{\rm L}}\nonumber\\
\alpha_2&=&4 G\left[U-T\sin^2\varphi\right]\tan\varphi\left[
\ln\sqrt{x_1^2+(x_2\sin\varphi+x_3\cos\varphi)^2}
\right]_{x_3=-x_{\rm L}}^{x_3=x_{\rm S}-x_{\rm L}}.
\end{eqnarray}
In the limit where $(x_{\rm S}-x_{\rm L})\left|\cos\varphi\right|$ and 
$x_{\rm L}\left|\cos\varphi\right|$ are much larger than $|x_1|$ and 
$\left|x_2\sin\varphi\right|$, we get 
\begin{eqnarray}\label{deftilt}
\alpha_1&\simeq&4\pi G\left[U\cos\varphi
+(U-T)\sin\varphi\tan\varphi\right]\nonumber\\
\alpha_2&\simeq&4 GU\tan\varphi\ln\frac{x_{\rm S}-x_{\rm L}}{x_{\rm L}}.
\end{eqnarray}
This has  to be compared with  the standard result for  a Goto--Nambu string
for              which              $\alpha=4\pi              GU\cos\varphi$
\cite{vilenkin,vilenkin84,hindmarsh2,hindmarsh90}.   Now, as pointed  out by
Peter \cite{peter94}  in the case of  a string perpendicular to  the line of
sight,  there are two  origins to  the deflection:  the deficit  angle (term
proportional to  $U+T$) and a contribution from  the curvature (proportional
to  $U-T$).   One  can   understand  such  a   result  by   decomposing  the
stress--energy   tensor  (\ref{cs})   as   $2{\cal  F}_{\mu\nu}=2U\times{\rm
diag}(0,1,0,1)+    (U-T)\times{\rm   diag}(1,-1,1,-1)$,    i.e.     as   the
superposition  of  a  Goto--Nambu   string  and  a  linear  distribution  of
non--relativistic matter  of density $\rho\equiv U-T$ per  unit length. Then
it  is  straightforward  to  see  that  the  bending  angle  of  the  second
contribution depends  only on  the projected mass  per unit length  and thus
becomes larger  by a factor  $1/\cos\varphi$ as found in  (\ref{deftilt}). A
consequence  of  this  result  is  that,  for  general  cosmic  strings  not
perpendicular to  the line of sight,  one expects to  have larger deflection
than for a Goto--Nambu string.

\begin{figure} 
\centerline{
\epsfig{figure=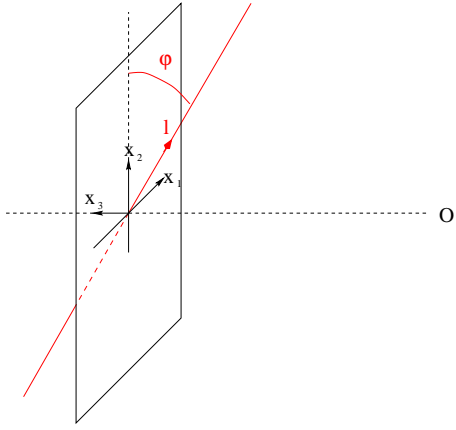,width=6cm}}
\caption{Parameterization of a static straight tilted cosmic string.}
\label{tilt} 
\end{figure}

In  this  case,  the  accuracy   of  the  thin  lens  approximation  can  be
investigated. For that purpose, we remind that the shear is given by
\begin{equation}
\left(\begin{array}{c}\gamma_1\\ \gamma_2\end{array}\right)=
2G\int_{-x_{\rm L}}^{x_{\rm S}-x_{\rm L}}\frac{(x_{\rm L}+x_3)(x_{\rm
S}-x_{\rm L}-x_3)}{x_{\rm S}}
\left(\begin{array}{c}\frac{1}{2}\left[\partial_2^2-\partial_1^2\right] \\ 
\partial_{1}\partial_{2}
\end{array}\right)\int_{-\infty}^\infty \frac{\left[U(\ell)-T(\ell)
\sin^2\varphi\right]\dd\ell}{\sqrt{x_1^2+(x_2-\ell\cos\varphi)^2+
(x_3-\ell\sin\varphi)^2}}.
\end{equation}
Due to the  derivatives with respect to  $x_1$ and $x_2$, it is  easy to see
that the  integral over $\ell$ is  peaked around $x_3\sim\ell\sin\varphi\sim
x_2\tan\varphi$. Thus,  on a  field of width  $x_2=x_{\rm S}  \theta_2$, the
variation of geometric factor is bounded by
$$
\frac{\delta\left|(x_{\rm L}-x_3)(x_{\rm S}-x_{\rm L}-x_3)\right|}{x_{\rm L}(x_{\rm S}-x_{\rm
L})}
\simeq 
\frac{\left(1+2\frac{x_{\rm L}}{x_{\rm S}}\right)}
{x_{\rm L}/x_{\rm S}\left(1-x_{\rm L}/x_{\rm S}\right)}
\left|\tan\varphi\right|\theta_2<\frac{3}{x_{\rm L}/x_{\rm S}\left(1-
x_{\rm L}/x_{\rm S}\right)}\left|\tan\varphi\right|\theta_2,
$$ 
from which we deduce that  since $\theta_2\ll1$, the thin lens approximation
is  still  very  good  for  tilted  string  with  a  tilt  not  larger  than
$\varphi=\pi/4$ say (see figure~\ref{err}  for a numerical estimation of the
relative error).

\begin{figure} 
\centerline{
\epsfig{figure=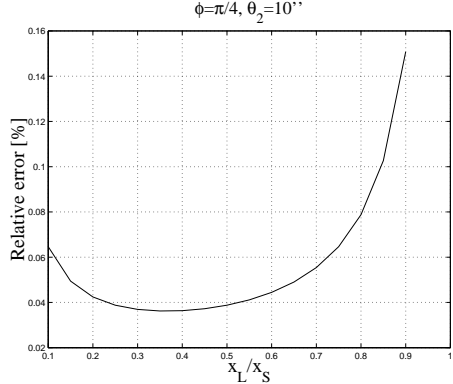,width=6cm}}
\caption{Relative error  on the  geometric factor with  respect to  the thin
lens  result  for   a  string  with  a  tilt   $\varphi=\pi/4$  on  a  field
$\theta_2=10''$.        The      relative       error       scales      like
$(\theta_2/10'')\times\tan\varphi$.}
\label{err} 
\end{figure}

\subsection{Discussion}

%In the general case of an extended lens, one cannot assume that the
%deflector is optically thin and thus the computation of the deflection
%is not sufficient and one has to compute directly the amplification
%matrix. 

In this section, we have shown that the deflecting potential of any extended
lens  with relativistic  motion  is obtained  by  considering the  projected
energy density on  the photon past light cone. This implies,  in the case of
cosmic strings, that the convergence $\kappa$ vanishes everywhere but on the
string projection.

We then studied  the case of a loop oscillating in  a plane perpendicular to
the line of  sight and show that the  equation of motion of the  loop can be
used  to  integrate  the  deflecting  potential.  We  found  that  a  photon
propagating  inside  such  a  circular  loop was  not  deflected  and  those
propagating outside were  deflected as if the loop was  a point mass object.
This  generalizes the  result by  de Laix  and Vachaspati  \cite{delaix2} to
strings with any  equation of state and shows how the  equation of motion of
the loop enables to factorize the integrated deflecting potential. This lets
us conjecture  that this result will  be valid whatever the  geometry of the
loop,   the   geometric  factor   $J$   being   different   for  each   loop
geometry. We finished by discussing the case of static tilted cosmic strings
to  emphasize  that,  for  non--Goto--Nambu  strings, there  can  be  larger
deflections and we  also discussed on this example the  validity of the thin
lens approximation for strings.

%%%%%%%%%%%%%%%%%%%%%%%%%%%%%%%%%%%%%%%%%%%%%%%%%%%%%%%%%%%%%%%%%%%
\section{Phenomenology of a deformation field with $\kappa=0$}\label{IV}

As seen in the previous section, we expect the deformation field of a cosmic
string to be  such that $\kappa=0$. Indeed, in  (\ref{kappa}) we only showed
that $\kappa$  vanishes in all directions  such that the  light ray arriving
with  this direction  does  not  intersect the  string  worldsheet. In  this
section, we are mainly interested in objects which do not overlap the string
such that  their deformation  is the one  with a $\kappa=0$  field.  Another
restriction of this  study is that we assume that  the caracteristic size of
the object  is smaller than the  caracteristic size of the  variaiton of the
shear  $\gamma$; thus  we donnot  consider galaxies  lying in  the immediate
neighborhood  of the  string (see  the companion  article~\cite{bu2}  for an
illustration). With these restrictions, we can consider that the deformation
field has a  zero convergence and the  goal of this section is  to study the
main phenomenological properties of such a field. The two kinds of questions
we would like to answer are:
\begin{itemize}
\item[(Q1)] Given  a source of  surface $S^{\rm S}$ and  ellipticity $e^{\rm
S}$, what can we say about the morphologies of all its possible images?
\item[(Q2)] Given two  objects, how can we know that they  are the images of
the same source? This question  reduces to study the allowed morphologies of
the sources that have the same images.
\end{itemize}
We start by  setting the general framework and  then study respectively (Q1)
and (Q2). This study is a first step toward the discrimination between pairs
of lensed sources by a cosmic string and fake lenses \cite{cowie,emhu}. This
study is made within the assumption that the shear variations over observed
background images is small. It may actually be a severe limitation for such
an approach if the string energy density is small.

\subsection{Describing the morphology of a cosmic object}

To  any object  of elliptic  shape such  as a  galaxy or  a cluster,  we can
associate a positive definite symmetric matrix $M_{ab}$ describing its shape
as
\begin{equation}
X^t M X \leq 1
\end{equation}
where $X^t=(x_1,x_2)$.
This matrix can be diagonalised as
\begin{equation}\label{101}
M=P^t(\theta)\left(\begin{array}{cc}
\lambda_+&0\\ 0&\lambda_-\end{array}\right)
P(\theta)
\end{equation}
where $P$ is a rotation matrix defined by
\begin{equation}
P(\theta)\equiv\left(\begin{array}{cc}
\cos\theta&\sin\theta\\ -\sin\theta&\cos\theta\end{array}\right)
\end{equation}
and a subscript $t$ denotes the transposition.  $\lambda_-\leq\lambda_+$ are
the two positive eigenvalues of $M$  and $\theta$ is an angle describing the
orientation of its principal axis  with respect to the basis $n^\mu_a$. Thus
any  object  can  be   characterized  by  the  set  ($\theta$,  $\lambda_-$,
$\lambda_+$) from which we can define the surface $S$ and ellipticity $e$ of
the object respectively as
\begin{eqnarray}
S(M)&\equiv&\hbox{det}(M)=\lambda_+\lambda_-,\\
e(M)&\equiv&\frac{\lambda_+-\lambda_-}{\lambda_++\lambda_-}=
        \sqrt{1-4\frac{\hbox{det}(M)}{[\hbox{tr}(M)]^2}}
\end{eqnarray}
and we also define $\epsilon$ as
\begin{equation}\label{def_epsilon}
\epsilon\equiv1-e^2=4\frac{\hbox{det}(M)}{[\hbox{tr}(M)]^2}.
\end{equation}
These definitions can indeed be inverted to get the two eigenvalues in terms
of $e$ and $S$ as
\begin{equation}
\lambda_\pm^2=S\left(\frac{1+e}{1-e}\right)^{\pm1}.
\end{equation}
Following Mellier \cite{mellier00}, the ellipticity must be bounded by
\begin{equation}\label{ellmax}
\epsilon\gtrsim0.5\Longleftrightarrow e\lesssim0.71.
\end{equation}

In the following, we will not be interested in the orientation of the
object and we then define the {\it shape} as being the set $(S,e)$. The
shape matrix $M^{\rm I}$ of any image can be related to the shape matrix
$M^{\rm S}$ of its associated source as (see e.g.~\cite{mellier00})
\begin{equation}
M^{\rm I}={\cal A}^{-1}M^{\rm S}{\cal A}^{-1}
\end{equation}
[this is valid only if we  consider that the carateristic size of the source
is smaller than the characteristic size associated with the variation of the
shear].  Decomposing   $M^{\rm  S}$   as  in  (\ref{101})   and  introducing
$\widetilde M^{\rm I}\equiv PM^{\rm I}  P^t$ which represents the same shape
as $M^{\rm I}$ after a rotation of $-\theta$, we obtain that
\begin{equation}
\widetilde M^{\rm I}=(P{\cal A}^{-1} P^t)
\left(\begin{array}{cc}\lambda_+&0\\ 0&\lambda_-\end{array}\right)
(P{\cal A}^{-1} P^t).
\end{equation}
Thus, $\widetilde M^{\rm  I}$ is the image of  the source $\widetilde M^{\rm
S}\equiv\left(\begin{array}{cc}\lambda_+&0\\   0&\lambda_-\end{array}\right)$
by the transformation,
\begin{equation}
\widetilde{\cal A}^{-1}\equiv P{\cal A}^{-1}P^t=\frac{1}{1-\widetilde\gamma^2}
\left(\begin{array}{cc}1+\widetilde\gamma_1&-\widetilde\gamma_2\\ 
-\widetilde\gamma_2&
1-\widetilde\gamma_1\end{array}\right)\quad
\hbox{with}\quad
\vec{\widetilde\gamma}=P(-2\theta)\vec\gamma.
\end{equation}
As  long  as  we  are  interested  only  on  the  shape  (i.e.  surface  and
ellipticity) of the sources and/or images,  we always can choose one of them
to be diagonal.

\subsection{Morphology of the images of a given source}

>From the previous analysis, we can  conclude that, if we are not interested
in the  relative orientation  of the source  and of  the image, we  can just
restrict the problem by considering the source and the transformation matrix
to be given by
\begin{equation}
M^{\rm S}=\left(\begin{array}{cc}\lambda_+&0\\ 0&\lambda_-\end{array}\right),
\quad
{\cal A}^{-1}=\frac{1}{1-\gamma^2}\left(\begin{array}{cc}1+\gamma_1&-\gamma_2\\ -\gamma_2&
1-\gamma_1\end{array}\right).
\end{equation}
Setting
\begin{equation}
\vec\gamma\equiv\gamma\left(\begin{array}{c}\cos\alpha\\ \sin\alpha\end{array}
\right)\quad\hbox{with}\quad \gamma\geq0\quad\hbox{and}\quad
\alpha\in [0,2\pi],
\end{equation} 
we can  easily show that the shape  $(S^{\rm I},e^{\rm I})$ of  the image is
related to the one of the source $(S^{\rm S},e^{\rm S}$) by
\begin{eqnarray}
S^{\rm I}&=&\frac{1}{(1-\gamma^2)^2}S^{\rm S}\label{SS0}\\
\epsilon^{\rm I}&=&\frac{(1-\gamma^2)^2}{(1+\gamma^2+2\gamma
e^{\rm S}\cos\alpha)^2}\epsilon^{\rm S}\label{eimage},
\end{eqnarray}
where  $\epsilon$ is  defined  in  (\ref{def_epsilon}).  In  the  case of  a
circular  source ($e^{\rm S}=0$)  we deduce  that, since  $(S^{\rm I}/S^{\rm
S},e^{\rm I})$ depends only on $\gamma$, (i) two images of same surface have
same ellipticity and  that (ii) all the  images lie on a curve  in the plane
$(S^{\rm  I}/S^{\rm S},e^{\rm  I})$.  In figure  \ref{fig1},  we depict  the
variation of  $S^{\rm I}$ and  $e^{\rm I}$ in  function of $\gamma$  and the
ensemble of the images of a circular source.
\begin{figure} 
\centerline{
\epsfig{figure=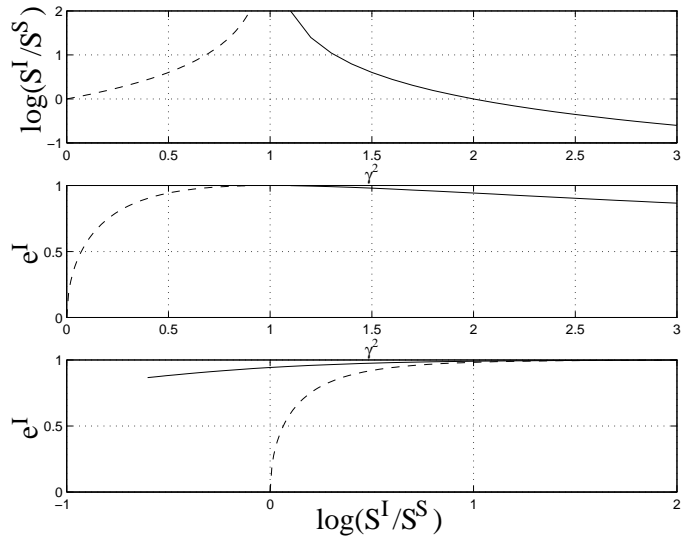,width=9cm}}
\caption{The  variation of $S^{\rm  I}/S^{\rm S}$  (top panel),  $e^{\rm I}$
(middle panel)  in function  of $\gamma$  and the set  of the  images (lower
panel) for a circular source ($e^{\rm S}=0$) of surface $S^{\rm S}$. The two
branches  represent respectively deformations  such that  $\gamma<1$ (dashed
line) and $\gamma>1$ (solid line).}
\label{fig1} 
\end{figure}
In  the  general   case  ($e^{\rm  S}\not=0$)  we  can   determine  all  the
morphologies of the images of a given source in the plane $(S^{\rm I}/S^{\rm
S},e^{\rm I})$. In figures \ref{fig2}, we depict these sets respectively for
$e^{\rm S}=0.25$ and $e^{\rm S}=0.5$.  We  note that all the curves have the
same asymptot  $(S^{\rm I}/S^{\rm  S})=\infty,e^{\rm I}=1$ which  is reached
when  $\gamma=1$,  i.e.   on  the  critical  line;  on   these  points,  the
amplification $\mu\equiv[{\rm  det}({\cal A})]^{-1}$ is  infinite.  Whatever
$e^{\rm  S}$,  there  exist  always  two circular  images  (i.e.  such  that
$\epsilon^{\rm I}=1$) obtained for
\begin{equation}\label{solgamma}
\gamma_\pm=\frac{e^{\rm S}}{1\mp\sqrt{\epsilon^{\rm S}}}\quad\alpha\equiv\pi\,[2\pi]
\end{equation}
(the condition on $\alpha$ is obtained from the requirement that $\gamma>0$;
there are  also two solutions  for $\alpha\equiv0\,[2\pi]$ but they  lead to
negative  values   of  $\gamma$).   Now,  since   from  (\ref{SS0})  $S^{\rm
I}_-/S^{\rm I}_+=(1-\gamma_-^2)^2/(1-\gamma_+^2)^2$, the measure of the area
of  two such circular  images enables  us (i)  to determine  the ellipticity
$e^{\rm S}$ of  their common source and (ii) the  shears $\gamma_\pm$ of the
two deformations. Indeed the bound  on the ellipticity (\ref{ellmax}) has to
be fulfilled by $e^{\rm  S}$ and this can be a test  to reject fake pairs of
images.

In  the more  general case  of a  pair of  non circular  images,  one cannot
reconstruct the ellipticity  of their source but we  can still conclude from
the  ratio  of  their   surfaces  if  they  correspond  to  transformations,
$\vec\gamma_1$ and $\vec\gamma_2$, that are both subcritical ($\gamma_{1}<1$
and  $\gamma_2<1$)  or where  one  is  critical  and the  other  subcritical
($\gamma_1<1$ and $\gamma_2>1$).

\begin{figure} 
\centerline{
\epsfig{figure=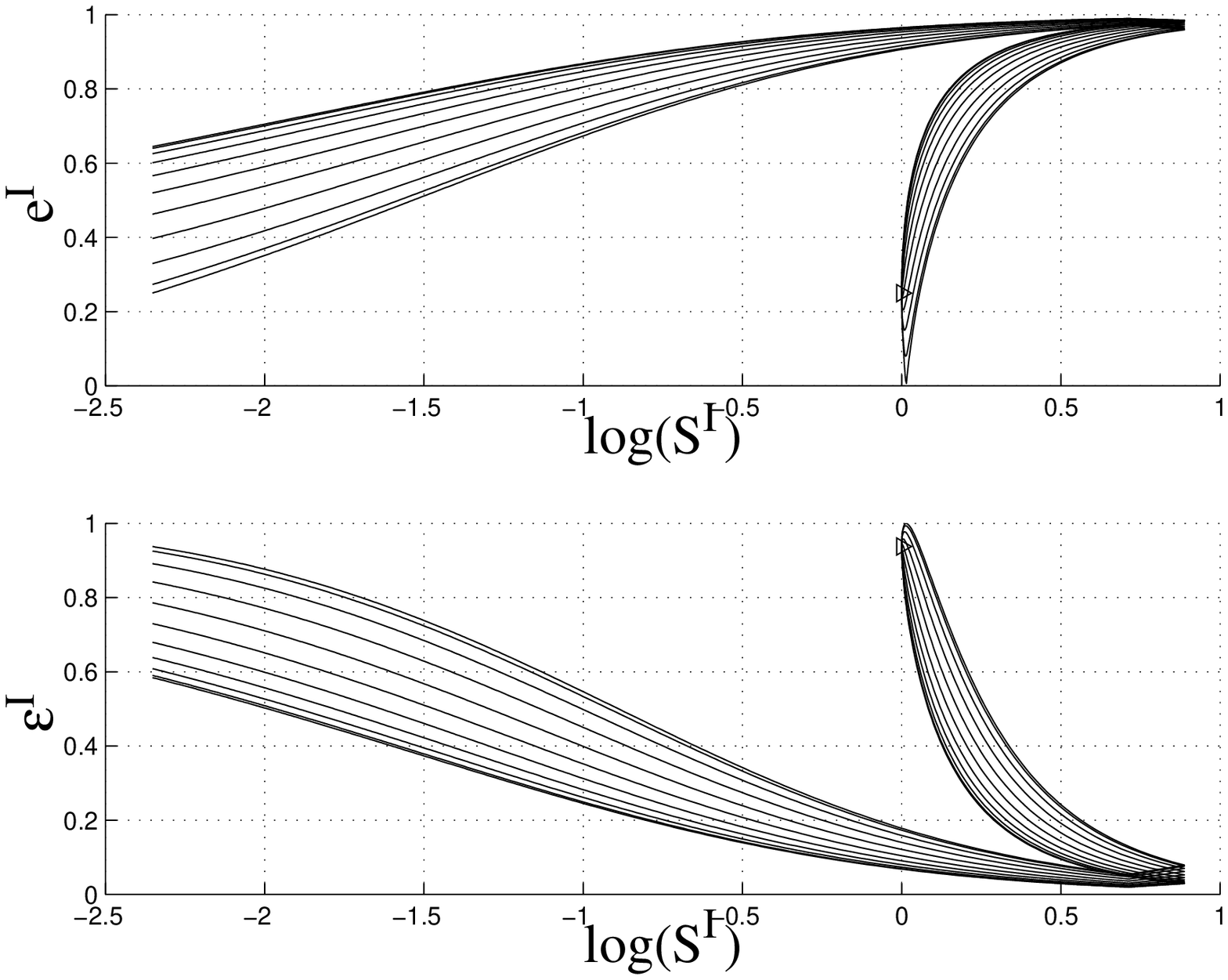,width=7cm}\epsfig{figure=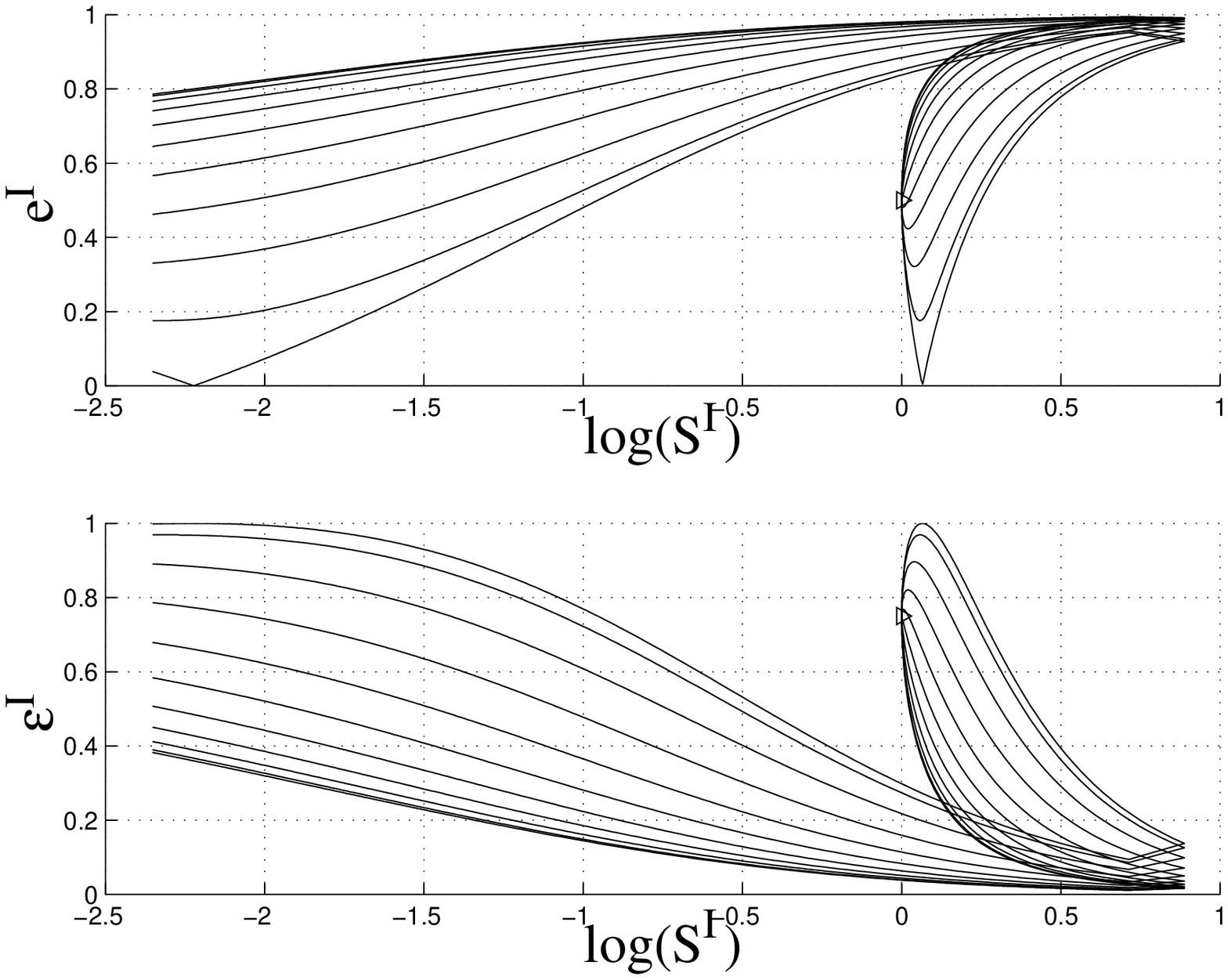,width=7cm}}
\caption{Set of all morpholohies ($(S^{\rm  I},e^{\rm I})$ of the image of a
source  of   ellipticity  $e^{\rm   S}=0.25$  [left]  and   $e^{\rm  S}=0.5$
[right]. As  explained in  the text,  we see (upper  right plot)  that there
always  exist two  circular  images, one  for  a subcritical  transformation
($\gamma=\gamma_-<1$)    and    one    for   a    critical    transformation
($\gamma=\gamma_+>1$)   [see    equation(\ref{solgamma})].   The   triangles
represent the images by the transformation with $\gamma=0$.}
\label{fig2} 
\end{figure}

For small deformations ($\gamma\ll1$), we have that
\begin{eqnarray}
S^{\rm I}/S^{\rm S}&\simeq&1+2\gamma^2+{\cal O}(\gamma^4)\\
\epsilon^{\rm I}/\epsilon^{\rm
S}&\simeq&1-4e^S\cos\alpha\gamma+4\left(3(e^S)^2\cos^2\alpha-1\right) 
\gamma^2+{\cal O}(\gamma^3)
\end{eqnarray}
so that the  images almost lie on a parabola  centered on $(S^{\rm I},e^{\rm
I})=(S^{\rm S},e^{\rm  S})$. In such  a weak field,  two images of  the same
object  will  have almost  the  same surface  but  can  have very  different
ellipticities.

\begin{figure} 
\centerline{
\epsfig{figure=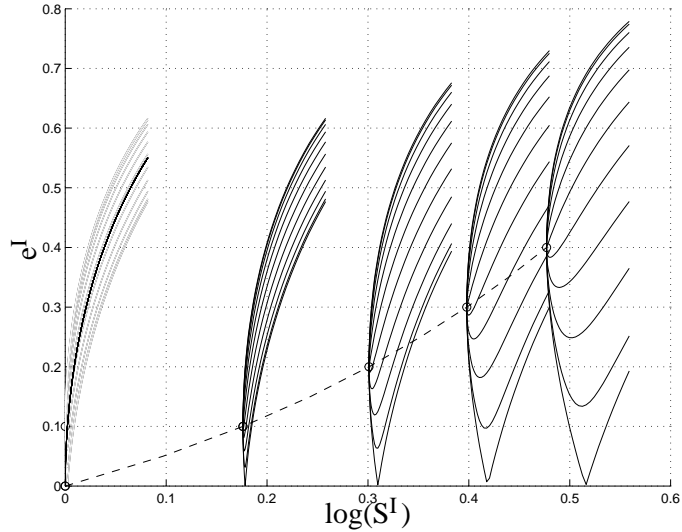,width=9cm}}
\caption{The images of the different sources of respective shape
(1,0), (1.5,0.1), (2,0.2), (2.5,0.3), (3,0.4) for a small
deformation $0<\gamma<0.3$. The circles represent the image
by the transformation $\gamma=0$.}
\label{fig4} 
\end{figure}

\subsection{Morphology of possible sources of a given image}

We now  ask the  reverse question: given  an image $(S^{\rm  I},e^{\rm I})$,
from which  set of sources can  it be the image?   Following the description
and notations of the two previous sections, we now set
\begin{equation}
M^{\rm I}=\left(\begin{array}{cc}
\lambda_+&0\\ 0&\lambda_-\end{array}\right),\quad
{\cal A}=\left(\begin{array}{cc}
1-\gamma_1&\gamma_2\\ \gamma_2&1+\gamma_1\end{array}\right)
\end{equation}
from which we deduce that, since $M^{\rm S}={\cal A}M^{\rm I}{\cal A}$,
the shape of the source is related to the one of its images by
\begin{eqnarray}
S^{\rm S}&=&(1-\gamma^2)^2S^{\rm I}\\
\epsilon^{\rm S}&=&\frac{(1-\gamma^2)^2}{(1+\gamma^2-2\gamma 
e^{\rm I}\cos\alpha)^2}\epsilon^{\rm I}.
\end{eqnarray}
As a  first exercise,  we depict on  figure~\ref{fig5} the  sources $(S^{\rm
S},e^{\rm  S})$ of  a  circular image  ($e^{\rm  I}=0$). A  priories on  the
ellipticity  of  the sources  (\ref{ellmax})  and  on  the strength  of  the
deformation $\gamma$ may enable us  to extract from such a plot informations
about the source of a circular image.

In  the general case  where $e^{\rm  I}\not=0$, we  can reconstruct  all the
morphologies  of the  source that  can  lead to  the observed  image. As  an
example  we depict  such sets  in the  two cases  where $e^{\rm  I}=0.5$ and
$e^{\rm I}=0.25$ respectively on  figures~\ref{fig6} and \ref{fig7}. We note
that all  the curves  pass through the  point $(S^{\rm  S},e^{\rm S})=(0,1)$
which is reaches when $\gamma=1$, i.e. on the critical line.

Again, one  can sort out that  any object can  be the image of  two circular
sources obtained by the two transformations
\begin{equation}
\gamma_\pm=\frac{e^{\rm I}}{1\mp\sqrt{\epsilon^{\rm I}}}\quad\alpha\equiv0\,[2\pi]
\end{equation}
and the measure of $(S^{\rm I},e^{\rm I})$ enables to determine $\gamma_\pm$
and the two surfaces $S_\pm^{\rm S}$.

\begin{figure} 
\centerline{
\epsfig{figure=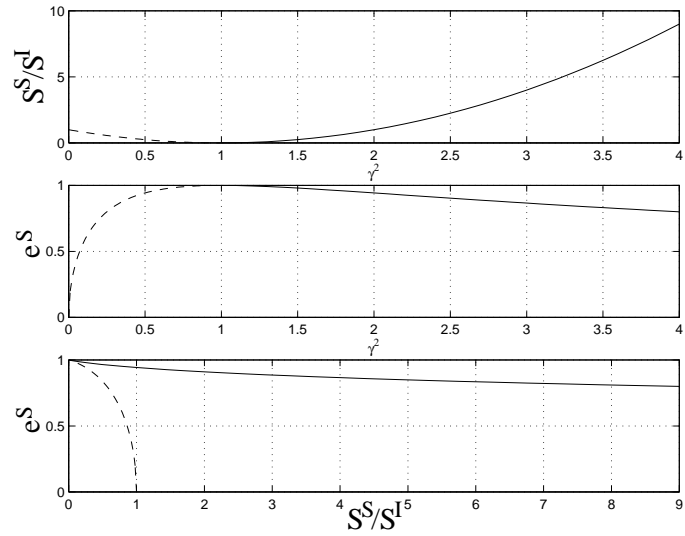,width=9cm}}
\caption{Sources of a circular image. We show the variation
of $S^{\rm S}/S^{\rm I}$ (upper panel) and of $e^{\rm S}$ (middle panel) with respect to the
strength of the deformation $\gamma$. The two branches correspond
respectively to transformations such that $\gamma<1$ (dash line) and
$\gamma>1$ (solid line).}
\label{fig5} 
\end{figure}

\begin{figure} 
\centerline{
\epsfig{figure=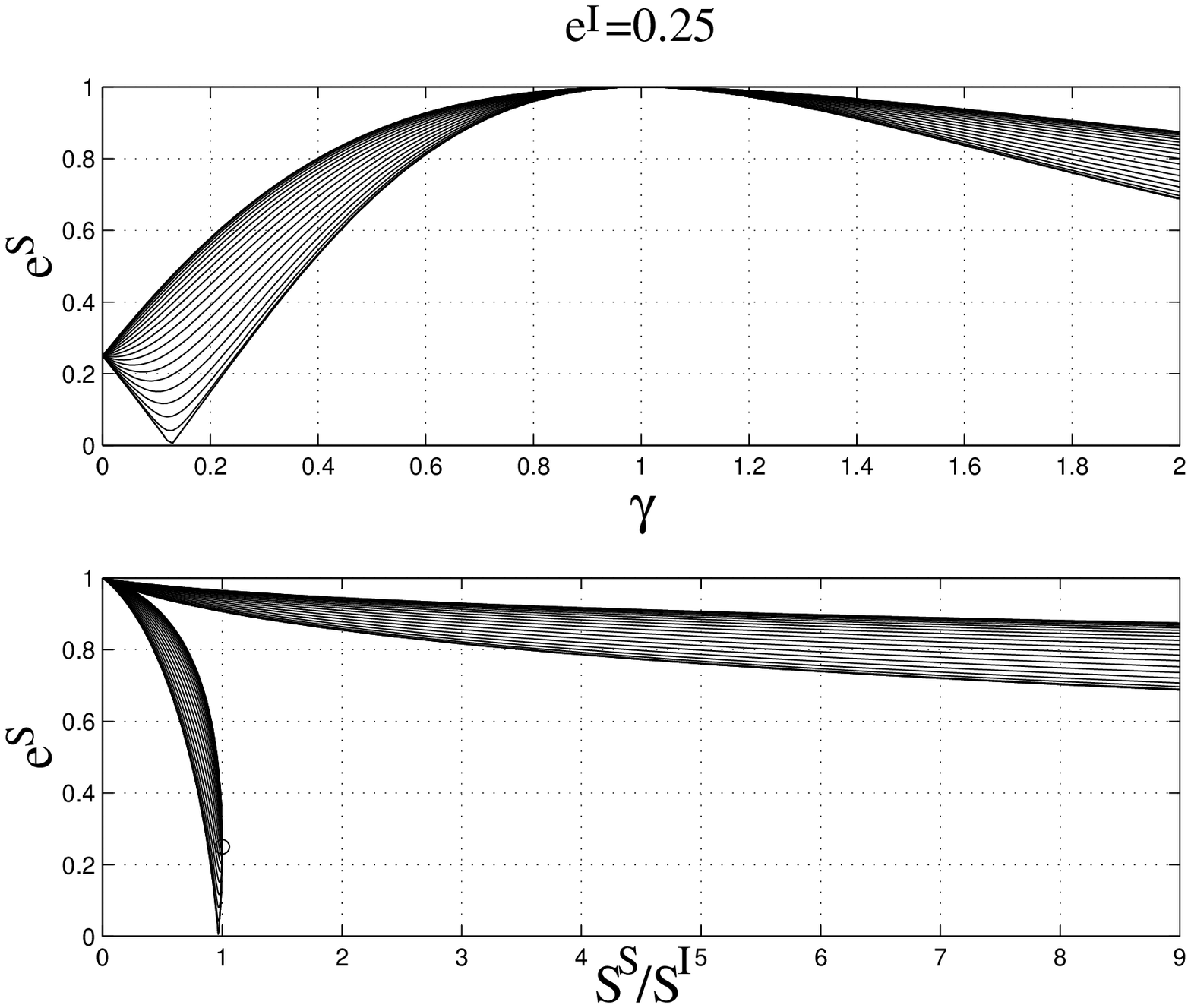,width=7cm}
\epsfig{figure=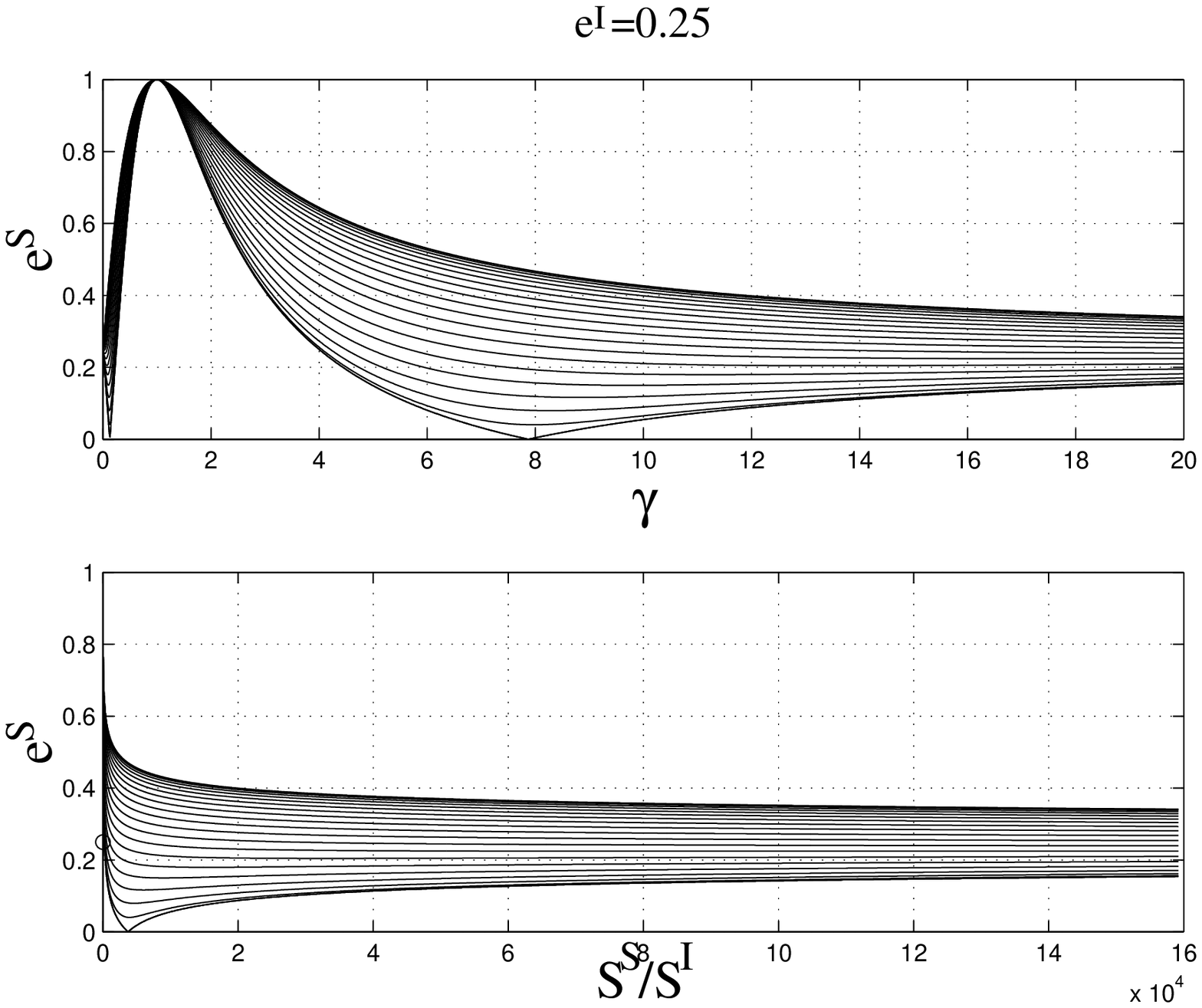,width=7cm}}
\caption{Variation of the ellipticity of the source of an image of
ellipticity $e^{\rm I}=0.25$. The circles represent the deformation with
$\gamma=0$.}
\label{fig6} 
\end{figure}

\begin{figure} 
\centerline{
\epsfig{figure=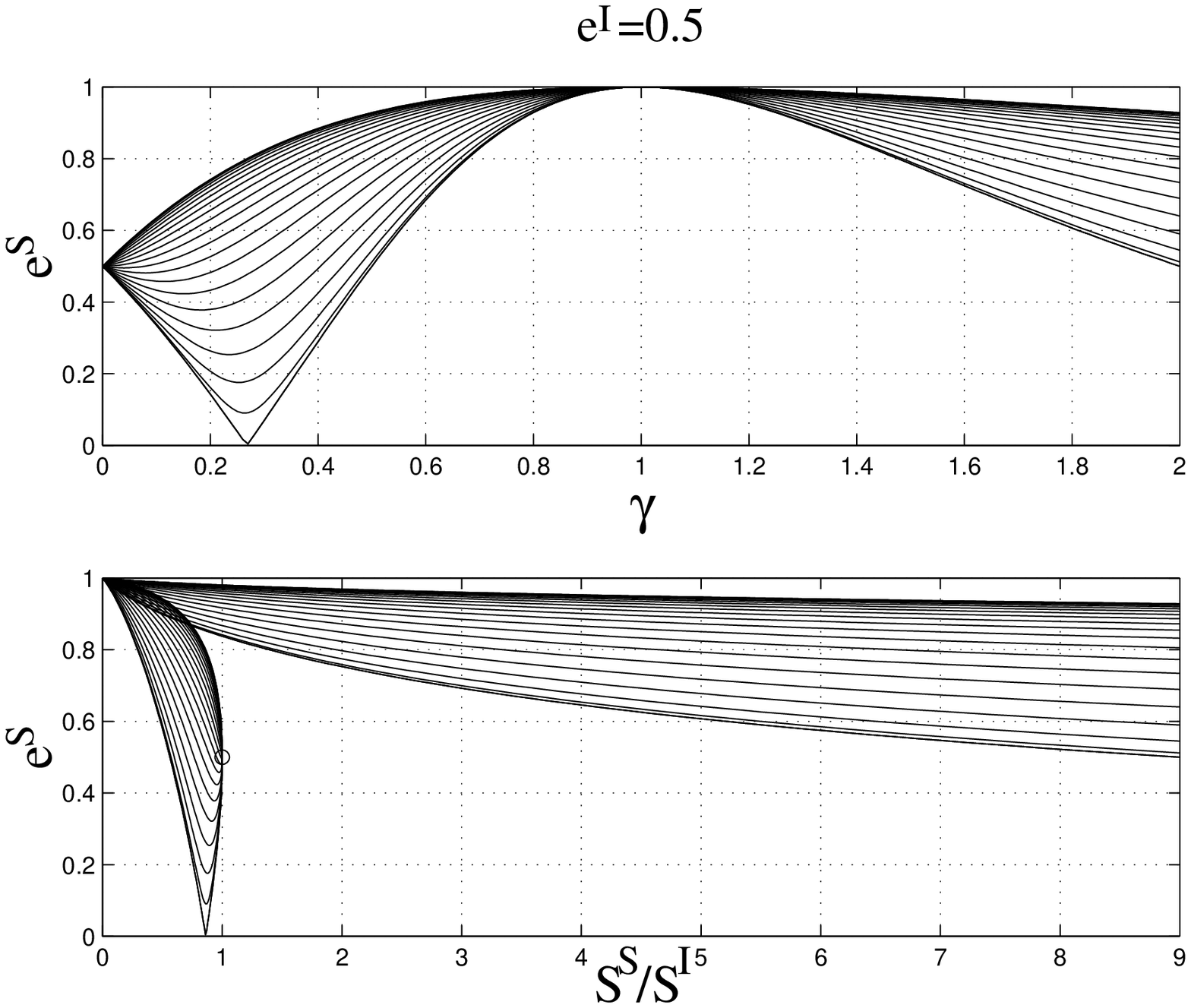,width=7cm}
\epsfig{figure=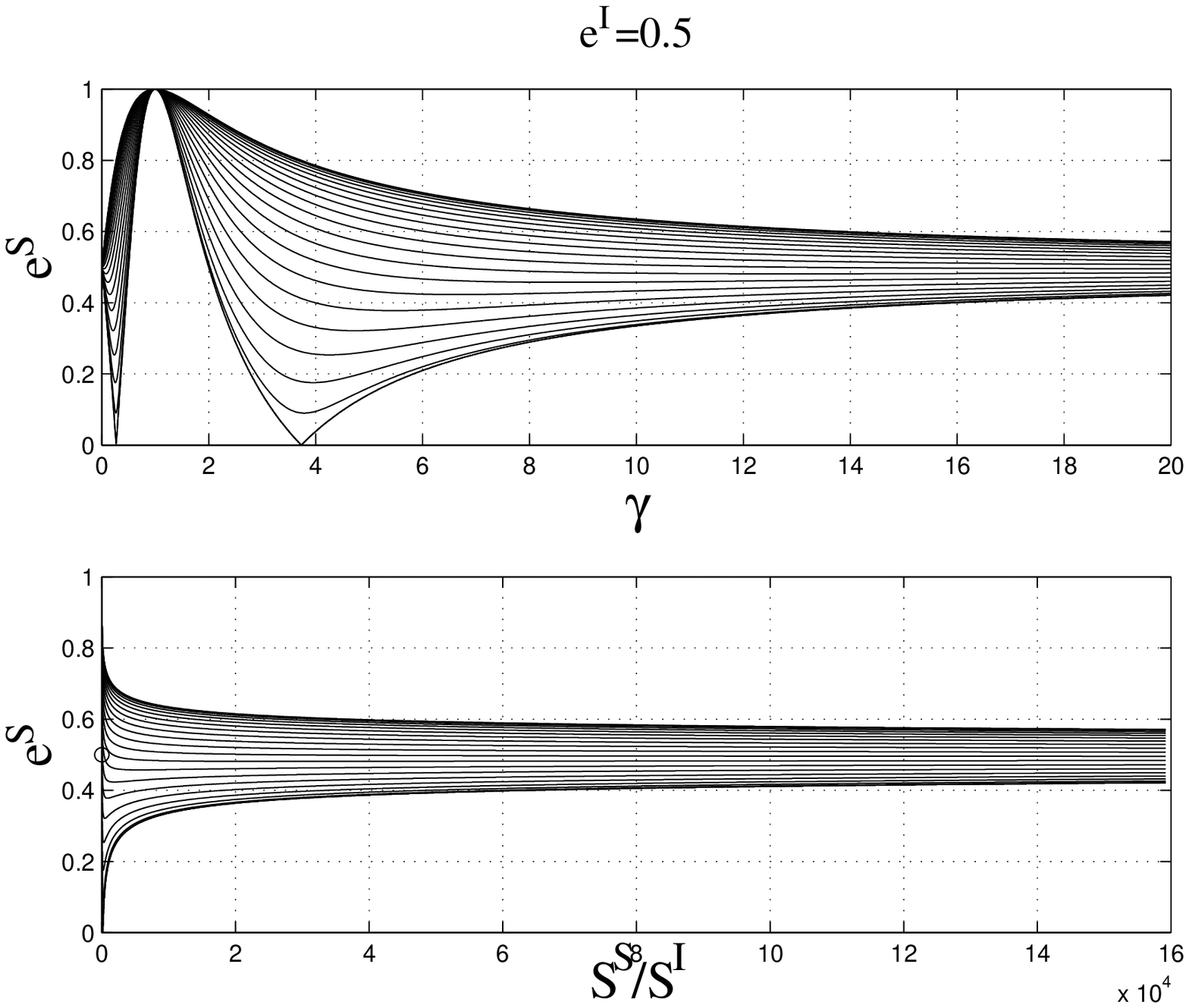,width=7cm}}
\caption{Variation of the ellipticity of the source of an image of
ellipticity $e^{\rm I}=0.25$. The circles represent the deformation with
$\gamma=0$.}
\label{fig7} 
\end{figure}

If  one measures the  shape of  two images  $(S^{\rm I}_1,e^{\rm  I}_1)$ and
$(S^{\rm I}_2,e^{\rm  I}_2)$ one  can reconstruct, as  in figures~\ref{fig6}
and \ref{fig7},  the set  of morphologies of  their possible  sources. Given
bounds on $e^{\rm S} $, as in (\ref{ellmax}), and on $\gamma$ one can figure
out graphically  if these two  observations are likely  to be images  of the
same object. Indeed for very  weak deformations ($\gamma\ll1$) one gets that
two images of the same objet must  have almost the same surface but can have
very different ellipticity (see  figure~\ref{fig8} where we have plotted the
source shapes of objects of different shape for small deformation).

\begin{figure} 
\centerline{
\epsfig{figure=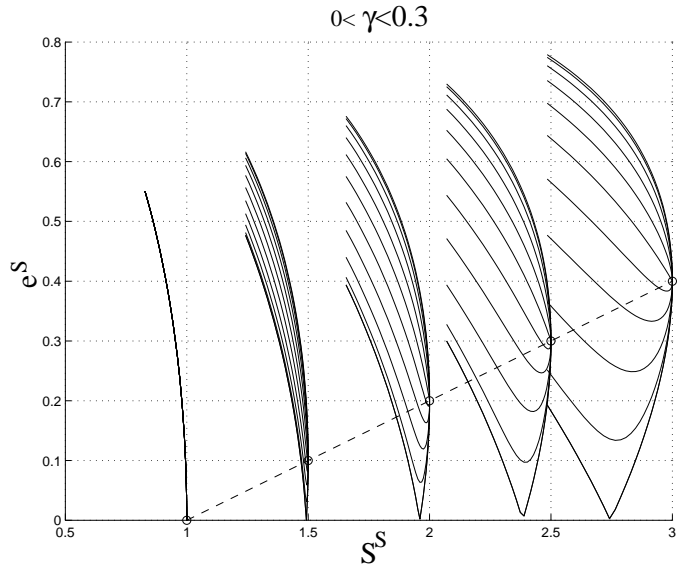,width=9cm}}
\caption{The source shapes of different images of respective shape
(1,0), (1.5,0.1), (2,0.2), (2.5,0.3), (3,0.4) for a small deformation
$0<\gamma<0.3$. The circles represent the deformation $\gamma=0$.}
\label{fig8} 
\end{figure}

%%%%%%%%%%%%%%%%%%%%%%%%%%%%%%%%%%%%%%%%%%%%%%%%%%%%%%%%%%%%%%%%%%%
\section{Conclusion}

We have considered  the general lensing properties of  objects such as cosmic
strings where relativistic motions  and non-trivial equation of state induce
metric  perturbations of all  sorts.  We  demonstrated that  the deformation
field of  a string system  on the image  plane is the same  as the one  of a
static linear distribution  of matter projected on the  photon trajectory. A
consequence of  this result  is that the  deformation field has  a vanishing
convergence   ($\kappa=0$)  everywhere   but  on   the  projection   of  the
intersection of the  observer past light cone and  the string worldsheet.  We
explicitly illustrate this result with  the case of a circular cosmic string
loop in a plane perpendicular to the line-of-sight for which we generalize the
results found  in \cite{delaix2}  to string with  any equation of  state. We
also showed that (i) the fact that the deformation field outside the loop is
equivalent to the one obtained by a  massive point and (ii) that a light ray
passing inside the loop is not deflected are due to the energy conservation
of the string.

We also paid  attention to the validity of thin  lens approximation for this
unusual  lens   system.   This  approximation is discussed in detail through
the case of a  static tilted straight cosmic string. It lead
us  to point  out that  for string  with a  general equation  of  state, the
deflection may be  more important than for a  Goto--Nambu string, this being
understood by  the fact that the  stress--energy tensor of  a general cosmic
string can always be decomposed as the superposition of a Goto--Nambu string
and  a lineic distribution  of non--relativistic  matter. The  deflection is
then due  to the  combined effect  of the deficit  angle of  the Goto--Nambu
string and of the curvature induced by the lineic distribution of matter.

We also studied  general phenomenological  consequences  of deformation
fields with zero convergence on  multiple image systems, the main goal being
to be able to  assess if two images are likely to form  an image pair of the
same source.  For that purpose, we derived  all the image shapes  of a given
source as well as all the source  shapes of a given image. These results may
serve as a groundwork for the elaboration of string detection strategies.

We are aware that this latter part  is limited by the fact that we only took
advantage  of the  zero-convergence property.   In practical  more intricate
distortion properties of the images are  likely to be useful. This is one of
the  aim   of  the  companion  paper  \cite{bu2}   where  more  quantitative
phenomenological  properties  are presented  that  take  into account  local
string energy fluctuations.

%%%%%%%%%%%%%%%%%%%%%%%%%%%%%%%%%%%%%%%%%%%%%%%%%%%%%%%%%%%%%%%%%%%
\section*{Acknowledgements}

We would like to thank Ruth Durrer, Yannick Mellier, Patrick
Peter and Albert Stebbins for discussions on the subject
and the anonymous referee for his numerous comments and corrections.

%%%%%%%%%%%%%%%%%%%%%%%%%%%%%%%%%%%%%%%%%%%%%%%%%%%%%%%%%%%%%%%%%%%

%%%%%%%%%%%%%%%%%%%%%%%%%%%%%%%%%%%%%%%%%%%%%%%%%%%%%%%%%%%%%%%%%%% 

\end{document}